# POPPER, KUHN, LAKATOS AND AIM-ORIENTED EMPIRICISM

Nicholas Maxwell
http://www.nick-maxwell.demon.co.uk
nicholas.maxwell@ucl.ac.uk

Abstract
   In this paper I argue that aim-oriented empiricism (AOE), a conception of natural science that I have defended at some length elsewhere, is a kind of synthesis of the views of Popper, Kuhn and Lakatos, but is also an improvement over the views of all three. Whereas Popper's falsificationism protects metaphysical assumptions implicitly made by science from criticism, AOE exposes all such assumptions to sustained criticism, and furthermore focuses criticism on those assumptions most likely to need revision if science is to make progress. Even though AOE is, in this way, more Popperian than Popper, it is also, in some respects, more like the views of Kuhn and Lakatos than falsificationism is. AOE is able, however, to solve problems which Kuhn's and Lakatos's views cannot solve.

1 Introduction
   In this paper I argue that aim-oriented empiricism (AOE), a conception of natural science that I have spelled out and defended at some length elsewhere,[1] is a kind of synthesis of the views of Popper, Kuhn and Lakatos, but is also an improvement over the views of all three.
   AOE stems from the observation that theoretical physics persistently accepts unified theories, even though endlessly many empirically more successful, but seriously disunified, *ad hoc* rivals can always be concocted. This persistent preference for and acceptance of unified theories, even against empirical considerations, means that physics makes a persistent untestable (metaphysical) assumption about the universe: the universe is such that no seriously disunified, *ad hoc* theory is true. Intellectual rigour demands that this substantial, influential, highly problematic and implicit assumption be made explicit, as a part of theoretical scientific knowledge, so that it can be critically assessed, so that alternative versions can be considered, in the hope that this will lead to an improved version of the assumption being developed and accepted. Physics is more rigorous when this implicit assumption is made explicit even though there is no justification for holding the assumption to be true. Indeed, it is above all when there is no such justification, and the assumption is substantial, influential, highly problematic, and all too likely to be false, that it becomes especially important to implement the above requirement for rigour, and make the implicit (and probably false) assumption explicit.
   Once it is conceded that physics does persistently assume that the universe is such that all seriously disunified theories are false, two fundamental problems immediately arise. What precisely ought this assumption to be interpreted to be asserting about the universe? Granted that the assumption is a pure conjecture, substantial and influential but bereft of any kind of justification, and thus all too likely in its current form to be false, how can rival versions of the assumption be rationally assessed, so that what is accepted by physics is improved?
   AOE is designed to solve, or help solve, these two problems. The basic idea is that we need to see physics (and science more generally) as making not one, but a hierarchy of assumptions concerning the unity, comprehensibility and knowability of the universe, the assumptions becoming less and less substantial as one goes up the hierarchy, and thus becoming more and more likely to be true: see diagram. The idea is that in this way we separate out what is most likely to be true, and not in need of revision, at and near the top of the hierarchy, from what is most likely to be false, and most in need of criticism and revision, near the bottom of the hierarchy. Evidence, at level 1, and assumptions high up in the hierarchy, are rather firmly accepted, as being most likely to be true (although still open to revision): this is then used to criticize, and to try to improve, theses at levels 2 and 3 (and perhaps 4), where falsity is most likely to be located.

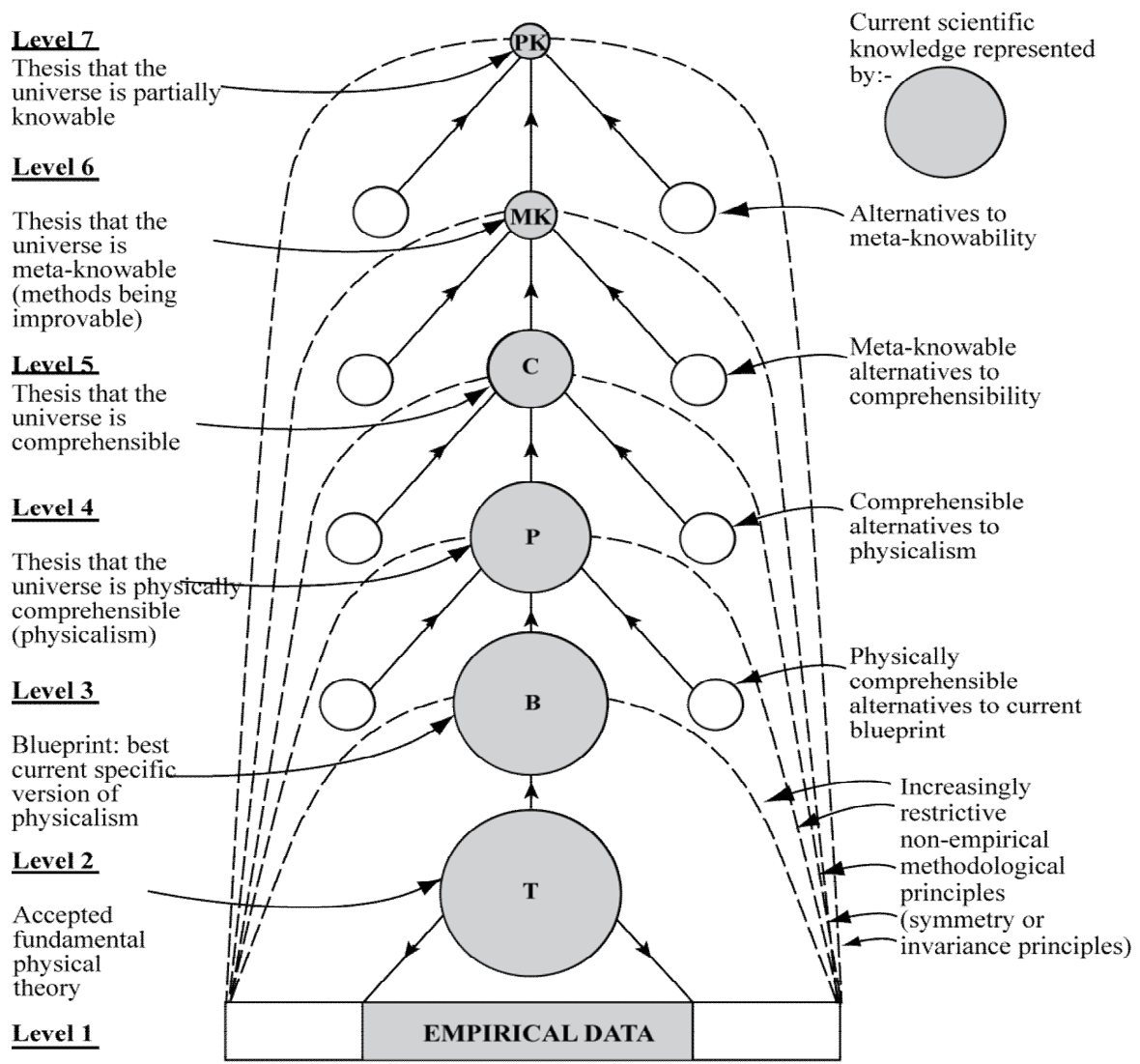

**Figure 1: Aim-Oriented Empiricism**

At the top there is the relatively insubstantial assumption that the universe is such that we can acquire some knowledge of our local circumstances. If this assumption is false, we will not be able to acquire knowledge whatever we assume. We are justified in accepting this assumption permanently as a part of our knowledge, even though we have no grounds for holding it to be true. As we descend the hierarchy, the assumptions become increasingly substantial and thus increasingly likely to be false. At level 5 there is the rather substantial assumption that the universe is comprehensible in some way or other, the universe being such that there is just one kind of explanation for all phenomena. At level 4 there is the more specific, and thus more substantial assumption that the universe is *physically* comprehensible, it being such that there is some yet-to-be-discovered, true, unified "theory of everything". At level 3 there is the even more specific, and thus even more substantial assumption that the universe is physically comprehensible in a more or less specific way, suggested by current accepted fundamental physical theories. Examples of assumptions made at this level, taken from the history of physics, include the following. The universe is made up of rigid corpuscles that interact by contact; it is made up of point-atoms that interact at a distance by means of rigid, spherically-symmetrical forces; it is made up of a unified field; it is made up of a unified quantum field; it is made up of quantum strings. Given the historical record of dramatically changing ideas at this level, and given the relatively highly specific and substantial character of successive assumptions made at this level, we can be reasonably confident that the best assumption available at any stage in the development of physics at this level will be false, and will need future revision. At level 2 there are the accepted fundamental theories of physics, currently general relativity and the standard model. Here, if anything, we can be even more confident that current theories are false, despite their immense empirical success. This confidence comes partly from the vast empirical content of these theories, and partly from the historical record. The greater the content of a proposition the more likely it is to be false; the fundamental theories of physics, general relativity and the standard model have such vast empirical content that this in itself almost guarantees falsity. And the historical record backs this up; Kepler's laws of planetary motion, and Galileo's laws of terrestrial motion are corrected by Newtonian theory, which is in turn corrected by special and general relativity; classical physics is corrected by quantum theory, in turn corrected by relativistic quantum theory, quantum field theory and the standard model. Each new theory in physics reveals that predecessors are false. Indeed, if the level 4 assumption of AOE is correct, then all current physical theories are false, since this assumption asserts that the true physical theory of everything is unified, and the totality of current fundamental physical theory, general relativity plus the standard model, is notoriously disunified.
Finally, at level 1 there are accepted empirical data, low level, corroborated, empirical laws.

In order to be acceptable, an assumption at any level from 6 to 3 must (as far as possible) be compatible with, and a special case of, the assumption above in the hierarchy; at the same time it must be (or promise to be) empirically fruitful in the sense that successive accepted physical theories increasingly successfully accord with (or exemplify) the assumption. At level 2, those physical theories are accepted which are sufficiently (a) empirically successful and (b) in accord with the best available assumption at level 3 (or level 4). Corresponding to each assumption, at any level from 7 to 3, there is a methodological principle, represented by sloping dotted lines in the diagram, requiring that theses lower down in the hierarchy are compatible with the given assumption.

When theoretical physics has completed its central task, and the true theory of everything, T, has been discovered, then T will (in principle) successfully predict all empirical phenomena at level 1, and will entail the assumption at level 3, which will in turn entail the assumption at level 4, and so on up the hierarchy. As it is, physics has not completed its task, T has not (yet) been discovered, and we are ignorant of the nature of the universe. This ignorance is reflected in clashes between theses at different levels of AOE. There are clashes between levels 1 and 2, 2 and 3, and 3 and 4. The attempt to resolve these clashes drives physics forward.

In seeking to resolve these clashes between levels, influences can go in both directions. Thus, given a clash between levels 1 and 2, this may lead to the modification, or replacement of the relevant theory at level 2; but, on the other hand, it may lead to the discovery that the relevant experimental result is not correct for any of a number of possible reasons, and needs to be modified. In general, however, such a clash leads to the rejection of the level 2 theory rather than the level 1 experimental result; the latter are held onto more firmly than the former, in part because experimental results have vastly less empirical content than theories, in part because of our confidence in the results of observation and direct experimental manipulation (especially after expert critical examination). Again, given a clash between levels 2 and 3, this may lead to the rejection of the relevant level 2 theory (because it is disunified, *ad hoc*, at odds with the current metaphysics of physics); but, on the other hand, it may lead to the rejection of the level 3 assumption and the adoption, instead, of a new assumption (as has happened a number of times in the history of physics, as we have seen). The rejection of the current level 3 assumption is likely to take place if the level 2 theory, which clashes with it, is highly successful empirically, and furthermore has the effect of increasing unity in the totality of fundamental physical theory overall, so that clashes between levels 2 and 4 are decreased. In general, however, clashes between levels 2 and 3 are resolved by the rejection or modification of theories at level 2 rather than the assumption at level 3, in part because of the vastly greater empirical content of level 2 theories, in part because of the empirical fruitfulness of the level 3 assumption (in the sense indicated above).

It is conceivable that the clash between level 2 theories and the level 4 assumption might lead to the revision of the latter rather than the former. This happened when Galileo rejected the then current level 4 assumption of Aristotelianism, and replaced it with the idea that "the book of nature is written in the language of mathematics" (an early precursor of our current level 4 assumption). The whole idea of AOE is, however, that as we go up the hierarchy of assumptions we are increasingly unlikely to encounter error, and the need for revision. The higher up we go, the more firmly assumptions are upheld, the more resistance there is to modification.

AOE is put forward as a framework which makes explicit metaphysical assumptions implicit in the manner in which physical theories are accepted and rejected, and which, at the same time, facilitates the critical assessment and improvement of these assumptions with the improvement of knowledge, criticism being concentrated where it is most needed, low down in the hierarchy. Within a framework of relatively insubstantial, unproblematic and permanent assumptions and methods (high up in the hierarchy), much more substantial, problematic assumptions and associated methods (low down in the hierarchy) can be revised and improved with improving theoretical knowledge. There is something like positive feedback between improving knowledge and improving (low-level) assumptions and methods - that is, knowledge-about-how-to-improve-knowledge. Science adapts its nature, its assumptions and methods, to what it discovers about the nature of the universe. This, I suggest, is the nub of scientific rationality, and the methodological key to the great success of modern science.

The above is intended to be an introductory account of AOE; further clarifications and details will emerge below when I come to expound AOE again during the course of arguing that the position can be construed to be a kind of synthesis of, and improvement over, the views of Popper, Kuhn and Lakatos.

In what follows I begin with Karl Popper, and argue that AOE can be seen to emerge as a result of modifying Popper's falsificationism[2] to remove defects inherent in that position. AOE does not, however, break with the spirit of Popper's work; far from committing the Popperian sin of "justificationism", AOE is even more Popperian than Popper, in that it is a conception of science which exposes more to effective criticism than falsificationism does. Falsificationism, in comparison, shields substantial, influential and problematic scientific assumptions from criticism within science. Whereas falsificationism fails to solve what may be called the "methodological" problem of induction, AOE successfully solves the problem. And, associated with that success, AOE also solves the problem of what it means to assert of a physical theory that it is "simple", "explanatory" or "unified", a problem which falsificationism fails to solve.

The conception of science expounded by Thomas Kuhn in his *The Structure of Scientific Revolutions* (1970) shares important elements with Popper's falsificationism. The big difference is that whereas Kuhn holds that "normal science" is an important, healthy and entirely rational (indeed, the most rational) part of science, Popper regards normal science as "dogmatic", the result of bad education and "indoctrination", something that is "a danger to science and, indeed, to our civilization" (Popper, 1970, p. 53). It is the apparent persistent dogmatism of normal science - the persistent retention of the current paradigm in the teeth of ostensible empirical refutations - that is so irrational, so unscientific, when viewed from a falsificationist perspective. AOE, however, though subjecting scientific assumptions to even greater critical scrutiny than Popper's falsificationism, turns out to have features which are, in some respects, closer to Kuhn than to Popper. For, according to AOE, substantial and influential metaphysical assumptions are persistently accepted as a part of scientific knowledge in a way which seems much closer to the way paradigms are accepted, according to Kuhn, during normal science, than to the way falsifiable theories are to be treated in science, according to Popper. AOE depicts science as, quite properly, proceeding in a way that is reminiscent, in important respects, of Kuhn's normal science, something that is anathema to Popper's falsificationism. At the same time, AOE is free of some of the serious defects inherent in Kuhn's conception of science. Even though AOE science mimics some aspects of Kuhnian normal science, it nevertheless entirely lacks the harmful dogmatism of this kind of science, and avoids problems that arise from Kuhn's insistence that successive paradigms are "incommensurable".

Imre Lakatos's "methodology of scientific research programmes"[3] was invented, specifically, to do justice both to Popper's insistence on the fundamental importance of subjecting scientific theories to persistent, ruthless attempted empirical refutation, and to Kuhn's insistence on the importance of preserving accepted paradigms from refutation, scientists, not paradigms, being under test when ostensible refutations arise. It is, like AOE, a kind synthesis of the ideas of Popper and Kuhn. Just as AOE incorporates elements of Popper and Kuhn, so too it incorporates elements of Lakatos's research programme methodology. At the same time, AOE is an improvement over Lakatos's view; it solves problems which Lakatos's view is unable to solve. Whereas Lakatos's view provides no means for the assessment of "hard cores" (Lakatos's "paradigms") other than by means of the empirical success and failure of the research programmes to which they give rise, AOE specifies a way in which "hard cores" (or their equivalent) can be rationally, but fallibly assessed, independent of the kind of empirical considerations to which Lakatos is restricted. This has important implications for the question of whether or not there is a rational method of discovery. It also has important implications for the *strength* of scientific method. For Lakatos, notoriously, scientific method could only decide which of two competing research programmes was the better long after the event, when one had proved to be vastly superior, empirically, to the other. "The owl of Minerva flies at dusk", as Lakatos put it, echoing Hegel. AOE provides a much more decisive methodology than Lakatos's, one which is able to deliver verdicts when they are needed, and not long after the event.

It may be thought that yet another critique of Popper, Kuhn and Lakatos is unnecessary, given the flood of literature that has appeared on the subject in the last 30 years or so: for an excellent recent survey article see Nola and Sankey (2000). My reply to this objection comes in two parts.

First, nowhere in this large body of critical literature can one find the particular line of criticism developed in the present paper. This line of criticism is, furthermore, especially fundamental and insightful in that it reveals, as other criticisms do not, what needs to be done radically to *improve* the views of Popper, Kuhn and Lakatos. Second, the improved view, namely AOE, that emerges from the criticism to be expounded here, has been entirely overlooked by the body of literature discussing and criticizing Popper, Kuhn and Lakatos. This is the decisive point. It is not enough merely to show that the views of Popper, Kuhn and Lakatos are defective. What really matters is to develop a view that overcomes these defects. That is what I set out to do here.

It is also true that, during the last 30 years, a substantial body of work has emerged on scientific method quite generally. I have in mind such publications as Holton (1973), Feyerabend (1978), Glymour (1980), van Fraassen (1980), Laudan (1984), Watkins (1984), Hooker (1987), Hull (1988), Howson and Urbach (1993), Kitcher (1993), Musgrave (1993), Dupré (1995), McAllister (1996), Cartwright (1999). In none of these works does one find the criticism of Popper, Kuhn and Lakatos, expressed below, or the synthesis, namely AOE, which emerges from this criticism. Furthermore, the methodological views developed in the works just cited all fall to the line of criticism deployed against Popper, Kuhn and Lakatos in the present paper. There is no space to develop this last point here: it is however spelled out in Maxwell (1998, ch. 2). One implication, then, of the present paper is that philosophy of science took a wrong turning around 1974 when it failed to take up the line of argument of this paper, an early version of which is to be found in Maxwell (1974).

2 Karl Popper

As everyone knows, Popper held that science proceeds by putting forward empirically falsifiable conjectures which are then subjected to severe attempts at falsification by means of observation and experiment. Scientific theories cannot be verified by experience, but they can be falsified. Once a theory is falsified, scientists have the task of developing a potentially better theory, even more falsifiable than its predecessor, at least as ostensibly empirically successful as its predecessor, and such that it is corroborated where its predecessor was falsified. In order to be accepted (tentatively) as a part of conjectural scientific knowledge a theory must (at least) be empirically falsifiable. Non-falsifiable, metaphysical theses are meaningful, and may influence the direction of scientific research. There can even be what Popper has called "metaphysical research programmes" - programmes of research "indispensable for science, although their character is that of metaphysical or speculative physics rather than of scientific physics ... more in the nature of myths, or of dreams, than of science" (Popper, 1982, p. 165). For Popper, metaphysical (that is, unfalsifiable) theses cannot be a part of (conjectural) scientific knowledge; such theses cannot help determine what is accepted and rejected as (conjectural) scientific knowledge, but they can influence ideas, choice of research aims and problems, in the context of scientific discovery. For further details see Popper (1959, 1963, 1983).

Popper defended two distinct versions of falsificationism which, echoing terminology of Maxwell (1998), I shall call bare and dressed falsificationism. According to bare falsificationism, defended in Popper (1959), only empirical considerations, and such things as the falsifiability of theories and degrees of falsifiability, decide what is to be accepted and rejected in science. According to dressed falsificationism, a new theory, in order to be acceptable, "should proceed from some *simple, new, and powerful, unifying idea* about some connection or relation (such as gravitational attraction) between hitherto unconnected things (such as planets and apples) or facts (such as inertial and gravitational mass) or new "theoretical entities" (such as field and particles)" (Popper, 1963, p. 241). This "requirement of simplicity" (as Popper calls it) is in addition to anything specified in Popper (1959). In his (1959), Popper does, it is true, demand of a theory that it should be as simple as possible, but Popper there identifies degree of simplicity of a theory with degree of falsifiability. (There is a second, related notion, but Popper makes it clear that if the two clash it is the falsifiability notion, just indicated, which takes priority: see page 130). Thus, in his (1959), in requiring of an acceptable theory that it should be as simple as possible, Popper is demanding no more than that it should be as falsifiable as possible. But Popper's "requirement of simplicity" of his (1963) is wholly in addition to falsifiability. A theory of high falsifiability may not "proceed from some simple, new, and powerful unifying idea", and vice versa. We thus have two versions of falsificationism before us: bare falsificationism of Popper's (1959), and dressed falsificationism of (1963, chapter 10), with the new "requirement of simplicity" added onto the (1959) doctrine.

I now give my argument for holding that neither doctrine is tenable. My argument is *not* that Popper fails to show how theories can be verified, or rendered probable; nor is my argument that Popper fails to show how scientific theories can be falsified, in that falsification requires the *verification* of a low-level falsifying hypothesis (which, according to Popper, is not possible).[4] There is nothing "justificationist", in other words, about my criticism. It amounts simply to this. Bare falsificationism fails dramatically to do justice to the way theories are selected in science (entirely independently of any question of verification, justification or falsification). Dressed falsificationism does better justice to scientific practice, but commits science to making substantial, influential and problematic assumptions that remain implicit, and cannot adequately be made explicit within science. Science pursued in accordance with dressed falsificationism is irrational, in other words, because it fails to implement the elementary, and quasi-Popperian, requirement for rationality that "assumptions that are substantial, influential, problematic and implicit need to be made explicit, so that they can be critically assessed and so that alternatives may be put forward and considered, in the hope that such assumptions can be improved" (Maxwell, 1998, p. 21). Dressed falsificationism fails, in other words, for good Popperian reasons: it fails to expose substantial, influential, problematic assumptions to *criticism* within science.

3 Refutation of Bare Falsificationism

Here, then, in a little more detail, is my refutation of bare falsificationism. Given any accepted physical theory, at any stage in the development of physics, however empirically successful (however highly corroborated) - Newtonian theory, say, or classical electrodynamics, quantum theory, general relativity, quantum electrodynamics, chromodynamics or the standard model - there will always be endlessly many rival falsifiable theories that can easily be formulated which will fit the available data just as well as the accepted theory. Taking Newtonian theory (NT) as an example of an accepted theory, here are two examples of rival theories. NT*: "Everything occurs as NT asserts, until the first second of 2100, when an inverse cube law of gravitation will abruptly hold". NT**: "Everything occurs as NT asserts, except for systems consisting of gold spheres, each having a mass of 1,000 tons, interacting with each other gravitationally in outer space, in a vacuum, within a spherical region of 10 miles: for these systems, Newton's law of gravitation is repulsive, not attractive". (For further examples and discussion, see Maxwell, 1998, pp. 47-54). It is easy to see that there are infinitely many such rivals to NT, just as empirically successful (at the moment) as NT. The predictions of NT may be represented as points in a multi-dimensional space, each point corresponding to some specific kind of system (there being infinitely many points). NT has only been verified (corroborated) for a minute region of this space. In order to concoct a (grossly *ad hoc*) rival to NT, just as empirically successful as NT, all we need do is identify some region in this space that includes no prediction of NT that has been verified, and then modify the laws of NT arbitrarily, for just that identified region.

The crucial question now is this: on what basis does bare falsificationism reject all these falsifiable but unfalsified rival theories? According to bare falsificationism, $T_2$ is to be accepted in preference to $T_1$ if $T_1$ has been falsified, $T_2$ has greater empirical content (is more falsifiable) than $T_1$, $T_2$ successfully predicts all that $T_1$ successfully predicts, $T_2$ successfully predicts the phenomena that falsified $T_1$, and $T_2$ successfully predicts new phenomena not predicted by $T_1$ (see Popper, 1959, pp. 81-84 and elsewhere). Given NT, it is a simple matter to concoct rival theories, of the above type, that satisfy all the above bare falsificationist requirements for being more acceptable than NT. Most accepted physical theories yield empirical predictions that clash with experiments, and thus are ostensibly falsified. We can always concoct new theories, in the way just indicated, doctored to yield the "correct" predictions. We can add on independently testable auxiliary postulates, thus ensuring that the new theory has greater empirical content than the old one. And no doubt this excess content will be corroborated. For details of how this can be done see Maxwell (1998, pp. 52-54). Such theories are, of course, grossly *ad hoc*, grossly "aberrant" as I have called them; but they satisfy Popper's (1959) requirements for being better theories than accepted physical theories.

It is worth noting that such "better" theories need not be quite as wildly *ad hoc* as the ones indicated above; sometimes such theories are actually put forward in the scientific literature, and yet are not taken seriously, even by their authors, let alone by the rest of the scientific community. An example is an *ad hoc* version of NT put forward by Maurice Levy in 1890, which combined in an *ad hoc* way two distinct modifications of Newton's law of gravitation, one based on the way Weber had proposed Coulomb's law should be modified, the other based on the way Riemann had proposed Coulomb's law should be modified: for details see North (1965). By 1890, NT had been refuted by observation of the precession of the perihelion of the orbit of Mercury; attempts to salvage NT by postulating an additional planet, Vulcan, had failed. Levy's theory successfully predicted all the success of NT, and in addition successfully predicted the observed orbit of Mercury, just that which refuted NT; in addition, of course, it made predictions different from NT for further Sun-Mercury type systems not yet observed. Despite this, Levy's theory was not taken seriously for a moment, not even by Levy himself. How can bare falsificationism recommend rejection of such *ad hoc* versions of NT when they satisfy all the requirements of bare falsificationism for being more acceptable theories? No adequate answer is forthcoming, and it is this which spells the downfall of bare falsificationism (as Popper may himself have realized when he put forward dressed falsificationism in his (1963), chapter 10).

Note, again, that this criticism of Popper has nothing justificational about it whatsoever: it simply points to the drastic failure of bare falsificationism to do justice to what actually goes on in physics.

It may be objected that *ad hoc* rivals to NT of the kind just considered are so silly, so crackpot, that they do not *deserve* to be taken seriously within physics.[5] This is of course correct. The crucial point, however, is that bare falsificationism ought to be able to deliver this verdict, and this it singularly fails to do. Bare falsificationism actually declares of appropriately concocted *ad hoc* rivals to NT that these are *better, more acceptable* than NT.

But can a criticism of Popper that appeals to such silly, crackpot theories be taken seriously? I have two replies to this question. First, not all the *ad hoc* or aberrant variants are entirely silly. Levy's theory is perhaps an example. There are degrees of *ad hocness*, from the utterly crackpot and absurd, to a degree of *ad hocness*, so slight, so questionable, in comparison, that the issue of whether the theory really is *ad hoc* or not may be hotly disputed by physicists themselves. (Such disputes arise especially during scientific revolutions.) This is an important point which will have a bearing on the argument of the next section. Second, it is, I submit, the very silliness of these crackpot theories that makes the above criticism of Popper so serious. If bare falsificationism favoured $T_1$ over $T_2$, while most scientists favoured $T_2$ over $T_1$, even though admitting that $T_1$ is nevertheless a good theory, almost as acceptable as $T_2$, bare falsificationism would not be in such trouble. What is lethal for bare falsificationism is that it declares $T_1$ to be better than $T_2$ in circumstances where scientists themselves (and all of us) can see that $T_2$ is vastly superior to $T_1$, $T_1$ being grossly *ad hoc*, aberrant, wholly crackpot and silly. Bare falsificationism favours theories that receive, and deserve, instant rejection: there could scarcely be a more decisive falsification of falsificationism than that.

4 Refutation of Dressed Falsificationism
   Having argued that Popper's (1959) bare falsificationism is untenable, I turn my attention now to Popper's (1963, chapter 10) doctrine of dressed falsificationism. As I have mentioned, this adds onto the (1959) doctrine Popper's new "requirement of simplicity (Popper, 1963, p. 241): see section 2 above.
   As long as there is no serious ambiguity as to what proceeding "from some simple, new, and powerful, unifying idea" means, it is at once clear that the new doctrine is able to exclude from science all the empirically successful but *ad hoc*, aberrant, crackpot, silly theories, of the kind discussed above. They do not proceed "from some simple...unifying idea", and are to be rejected on that account, whatever their empirical success may be, even if this empirical success is greater than accepted scientific theories.
   However, adopting Popper's new "principle of simplicity" as a basic methodological principle of science has the effect of permanently excluding from science all *ad hoc* theories that fail to satisfy the principle, however empirically successful such theories might be if considered. This amounts to assuming permanently that the universe is such that no *ad hoc* theory, that fails to satisfy Popper's principle of simplicity, is true. It amounts to accepting, as a permanent item of scientific knowledge, the substantial metaphysical thesis that the universe is non-*ad hoc*, in the sense that no theory that fails to satisfy Popper's principle of simplicity is true, however empirically successful it might turn out to be if considered. But this, of course, clashes with Popper's criterion of demarcation: that no unfalsifiable, metaphysical thesis is to be accepted as a part of scientific knowledge. If the demarcation principle is upheld, then the metaphysical thesis just indicated, asserting that the universe is non-*ad hoc*, remains implicit in the permanent adoption of Popper's principle of simplicity as a basic methodological principle of science. (And this is the way Popper himself seems to have conceived the matter: he says of metaphysical research programmes that they are "often held unconsciously", and "are implicit in the theories and in the attitudes and judgements of the scientists": (Popper, 1982, p. 161).) But in leaving the metaphysical thesis of non-*ad hocness* implicit in the methodological principle of simplicity, science violates an elementary requirement for rationality, already mentioned, according to which "assumptions that are substantial, influential, problematic and implicit need to be made explicit, so that they can be critically assessed and so that alternatives may be put forward and considered, in the hope that such assumptions can be improved" (Maxwell, 1998, p. 21). The non-*ad hoc* metaphysical assumption may, after all, be false. We may need to adopt a modified version of the assumption. It may be essential for the progress of science that this assumption is modified. Just this turns out to be the case, given certain formulations of the assumption, as we shall see below. In leaving the non-*ad hoc* metaphysical assumption implicit in the adoption of the methodological principle of simplicity, dressed falsificationism protects this substantial, influential and highly problematic assumption from *criticism*, from the active consideration of alternatives.[6]
   Dressed falsificationism fails, in other words, for good Popperian reasons: it is either inconsistent (in that the untestable, metaphysical thesis that the universe is non-*ad hoc* is held to be a part of conjectural scientific knowledge, in conflict with the principle of demarcation), or it irrationally protects an implicit, substantial assumption from explicit criticism within the intellectual domain of science.
   Here again, it should be noted, there is nothing justificationist about this criticism of Popper's dressed falsificationism. On the contrary, what the argument shows is that dressed falsificationism protects a substantial, influential, problematic but implicit assumption from criticism within science: Popper's doctrine fails for the good Popperian reason of restricting criticism.

It may be objected that adopting Popper's methodological principle of simplicity does *not* commit science to making a substantial metaphysical assumption about the universe - namely, that it is such that no falsifiable theory, however empirically successful, which fails to satisfy the principle, is true. But I do not see how such an objection can be valid. Suppose, instead of adopting Popper's principle, science adopted the principle: in order to be acceptable, a new physical theory must postulate that the universe is made up of atoms. This methodological principle is upheld in such a way that even though theories are available which postulate fields rather than atoms, and which are much more empirically successful than any atomic theory, nevertheless these rival field theories are all excluded from science. Would it not be clear that science, in adopting and implementing the methodological principle of atomicity in this way, is making the assumption that the universe is made up of atoms, whether this is acknowledged or not? How can this be denied? Just the same holds if science adopts and implements Popper's methodological principle of simplicity.

Popper might have tried to wriggle out of accepting this conclusion by pointing to the fact that he only declared that a new theory, in order to be acceptable, "should" proceed from some simple, unifying idea. It is desirable, but not essential, that new theories should satisfy this principle. The principle is relevant to the context of discovery, perhaps, but not to the context of acceptance and rejection. (It is a heuristic principle, not a methodological one.) But if Popper's doctrine is interpreted in this way, it immediately fails to overcome the objections spelled out in section 3 above. Either falsificationism adopts Popper's principle of simplicity as a *methodological* principle, or it does not. If it does, it encounters the objections just indicated; if it does not, it encounters the objections of section 3.

5 From Falsificationism to Aim-Oriented Empiricism

The conclusion to be drawn from the argument so far is that science is more rational, more intellectually rigorous, if it makes explicit, as a criticizable tenet of (conjectural) scientific knowledge, that substantial, influential and problematic metaphysical thesis which is implicit in the way physics persistently rejects *ad hoc* theories, however empirically successful they may be. At once two important new problems leap to our attention. What, precisely, does this metaphysical thesis assert? And on what grounds is it to be (conjecturally) accepted as a part of scientific knowledge? The conception of science which I uphold as a radical improvement over Popper's falsificationism, namely aim-oriented empiricism (AOE), is put forward as the solution to these two problems. I now expound AOE (in a little more detail than the introductory exposition of section 1) and indicate how it solves the two problems just mentioned; I indicate further how it solves the methodological problem of induction and the related problem of simplicity, and then consider possible objections.

As far as the first of the above two problems is concerned, a wide range of metaphysical theses are available. As I indicated in section 3 above, *ad hoc* theories range from the utterly crackpot and silly, to theories that are only somewhat lacking in simplicity or unity. At one extreme, we might adopt a metaphysical thesis that excludes only utterly silly theories; at the other extreme, we might adopt the thesis that the universe is physically comprehensible in the sense that it has a unified dynamic structure, some yet-to-be-discovered unified physical "theory of everything" being true - a thesis that I shall call "physicalism". We might even adopt some specific version of physicalism, which asserts that the underlying physical unity is of a specific type: it is made up of a unified field perhaps, or a quantum field, or empty topologically complex curved space-time, or a quantum string field. Other things being equal, the more specific the thesis (and thus the more it excludes) so the more likely it is to be false, whereas the more unspecific it is so the more likely it is to be true.

As far as the second of the above two problems is concerned, there are four considerations that we can appeal to, three wholly Popperian in spirit if not in the letter of Popperian doctrine.

(1) If some metaphysical thesis, M, is implicit in some scientific methodological practice, then science is more rigorous if M is made explicit, since this facilitates criticism of it, the consideration of alternatives.

(2) A metaphysical thesis may be such that its truth is a necessary condition for it to be possible for us to acquire knowledge: if so, accepting the thesis can only help, and cannot undermine, the pursuit of knowledge of truth.

(3) Given two rival metaphysical theses, $M_1$ and $M_2$, it may be the case that $M_1$ supports an empirical scientific research programme that has apparently met with far greater empirical success than any rival empirical research programme based on $M_2$: in this case we may favour $M_1$ over $M_2$, at least until $M_2$, or some third thesis, $M_3$, shows signs of supporting an even more empirically progressive research programme.

(4) $M_1$ may be preferred to $M_2$ on the grounds that it gives greater promise of supporting an empirically progressive research programme.

The arguments of sections 3 and 4 have established that physics must accept (conjecturally) some kind of metaphysical thesis of non-*ad hocness*, if crackpot theories are to be excluded: it makes sense to adopt that thesis which seems to be the most fruitful in promoting scientific progress. (To say that $M_1$ "supports" an empirically successful research programme is to say that the programme develops a succession of theories, each empirically more successful than its predecessors, in a Popperian sense, and each being closer to exemplifying, to being a precise, testable instantiation of, $M_1$ than its predecessors.)

Two difficulties arise, however, when one attempts to use (2) and (3) to select the best available metaphysical thesis from the infinitely many options available. First, as far as (2) is concerned, any thesis sufficiently substantial to exclude empirically successful crackpot theories from science is such that acquisition of knowledge might still be possible even if the thesis is false. On the other hand, any thesis such that its truth is necessary for knowledge to be acquired is much too insubstantial to exclude crackpot theories. Second, as far as (3) is concerned, given any metaphysical thesis, M, that supports a non-crackpot empirically progressive scientific research programme, we can mimic this with a crackpot M* that supports a crackpot empirically progressive research programme, with a series of crackpot theories, T1*, T2*, ..., these theories becoming progressively more and more empirically successful, and closer and closer to exemplifying M*.

These two difficulties can be overcome, however, if physics is construed as adopting a hierarchy of metaphysical conjectures concerning the comprehensibility and knowability of the universe, these conjectures becoming more and more insubstantial as one ascends the hierarchy, more and more likely to be true: see diagram. At level 7 there is the thesis that the universe is such that we can continue to acquire knowledge of our local circumstances, sufficient to make life possible. At level 6 there is the more substantial thesis that there is some rationally discoverable thesis about the nature of the universe which, if accepted, makes it possible progressively to improve methods for the improvement of knowledge. "Rationally discoverable", here, means at least that the thesis is not an arbitrary choice from infinitely many analogous theses. At level 5 we have the even more substantial thesis that the universe is *comprehensible* in some way or other, whether physically or in some other way. This thesis asserts that the universe is such that there is *something* (God, tribe of gods, cosmic goal, physical entity, cosmic programme or whatever), which exists everywhere in an unchanging form and which, in some sense, determines or is responsible for everything that changes (all change and diversity in the world in principle being explicable and understandable in terms of the underlying unchanging *something*).  A universe of this type deserves to be called "comprehensible" because it is such that everything that occurs, all change and diversity, can in principle be explained and understood as being the outcome of the operations of the one underlying *something*, present throughout all phenomena.  At level 4 we have the still more substantial thesis that the universe is *physically* comprehensible in some way or other (a thesis I shall call *physicalism*[7]).  This asserts that the universe is made up one unified self-interacting physical entity (or one kind of entity), all change and diversity being in principle explicable in terms of this entity.  What this amounts to is that the universe is such that some yet-to-be-discovered unified physical theory of everything is true.  At level 3, we have an even more substantial thesis, the best, currently available specific idea as to how the universe is physically comprehensible.  This asserts that everything is made of some specific kind of physical entity: corpuscle, point-particle, classical field, quantum field, convoluted space-time, string, or whatever.  Because the thesis at this level is so specific, it is almost bound to be false (even if the universe is physically comprehensible in some way or other).  Here, ideas evolve with evolving knowledge.  At level 2, we have our best fundamental physical theories, currently general relativity and the so-called standard model, and at level 1 we have empirical data (low level experimental laws).

The thesis at the top of the hierarchy, at level 7, is such that, if it is false, knowledge cannot be acquired whatever is assumed.  This thesis is, quite properly, accepted as a permanent part of scientific knowledge, in accordance with (2) above, since accepting it can only help, and cannot hinder, the acquisition of knowledge whatever the universe is like.

I have two arguments (appealing to (4) above) for the acceptance of the thesis of meta-knowability, at level 6.

(i) Granted that there is *some* kind of general feature of the universe which makes it possible to acquire knowledge of our local environment (as guaranteed by the thesis at level 7), it is reasonable to suppose that we do not know all that there is to be known about what the *nature* of this general feature is. It is reasonable to suppose, in other words, that we can improve our knowledge about the nature of this general feature, thus improving methods for the improvement of knowledge.  Not to suppose this is to assume, arrogantly, that we already know all that there is to be known about how to acquire new knowledge.  Granted that learning is possible (as guaranteed by the level 7 thesis), it is reasonable to suppose that, as we learn more about the world, we will learn more about how to learn.  Granted the level 7 thesis, in other words, meta-knowability is a reasonable conjecture.

(ii) Meta-knowability is too good a possibility, from the standpoint of the growth of knowledge, not to be accepted initially, the idea only being reluctantly abandoned if all attempts at improving methods for the improvement of knowledge fail.

These two arguments for accepting meta-knowability are, admittedly, weak.  It is crucial, however, that these two arguments make no appeal to the success of science, for a reason that will become apparent in a moment.

The thesis that the universe is comprehensible, at level 5 is accepted because no rival thesis, at that level, has been so fruitful in leading to empirically progressive research programmes.  It is hardly an exaggeration to say that all empirically successful research programmes into natural phenomena have been organized around the search for explanatory theories, of one kind or another.  Aberrant rivals to the thesis of comprehensibility, which might be construed as supporting aberrant empirically successful research programmes, are rejected because of incompatibility with the thesis of meta-knowability at level 6.  Such rival ideas are not "rationally discoverable" in that each constitutes an arbitrary choice from infinitely many equivalent rivals.

Physicalism at level 4 is accepted because it is by far the most empirically fruitful thesis at that level that is compatible with the thesis of comprehensibility, at level 5.

Since the scientific revolution of the 17th century, all new fundamental physical theories have enhanced overall unity of theoretical physics. Thus Newtonian theory (NT) unifies Galileo's laws of terrestrial motion and Kepler's laws of planetary motion (and much else besides). Maxwellian classical electrodynamics, (CEM), unifies electricity, magnetism and light (plus radio, infra red, ultra violet, X and gamma rays). Special relativity (SR) brings greater unity to CEM (in revealing that the way one divides up the electromagnetic field into the electric and magnetic fields depends on one's reference frame). SR is also a step towards unifying NT and CEM in that it transforms space and time so as to make CEM satisfy a basic principle fundamental to NT, namely the (restricted) principle of relativity. SR also brings about a unification of matter and energy, via the most famous equation of modern physics, $E = mc^2$, and partially unifies space and time into Minkowskian space-time. General relativity (GR) unifies space-time and gravitation, in that, according to GR, gravitation is no more than an effect of the curvature of space-time. Quantum theory (QM) and atomic theory unify a mass of phenomena having to do with the structure and properties of matter, and the way matter interacts with light. Quantum electrodynamics unifies QM, CEM and SR. Quantum electroweak theory unifies (partially) electromagnetism and the weak force. Quantum chromodynamics brings unity to hadron physics (via quarks) and brings unity to the eight kinds of gluon of the strong force. The standard model unifies to a considerable extent all known phenomena associated with fundamental particles and the forces between them (apart from gravitation). The theory unifies to some extent its two component quantum field theories in that both are locally gauge invariant (the symmetry group being U(1)XSU(2)XSU(3)). String theory, or M-theory, holds out the hope of unifying all phenomena. All these theories have been accepted because they progressively (a) *increase* the overall unity of theoretical physics and (b) *increase* the predictive power of physical theory, (a) being as important as (b). Physicalism is the key, persisting thesis of the entire research programme of theoretical physics since Galileo, and no obvious rival thesis, at that level of generality, can be substituted for physicalism in this research programme.

 It may be asked: But how can this succession of theories reinforce physicalism when the totality of physical theory has always, up till now, clashed with physicalism? The answer: If physicalism is true, then all physical theories that only unify a restricted range of phenomena, must be false. Granted the truth of physicalism, and granted that theoretical physics advances by putting forward theories of limited but ever increasing empirical scope, then it follows that physics will advance from one false theory to another (as it has done: see point 7 of section 6 below), all theories being false until a unified theory of everything is achieved (which just might be true). The successful pursuit of physicalism requires progressive *increase* in both empirical scope and unity of the totality of fundamental physical theory. It is just this which the history of physics, from Galileo to today, exemplifies - thus demonstrating the unique fruitfulness of physicalism.

 At level 3 that metaphysical thesis is accepted which is the best specific version of physicalism available, that seems to do the best justice to the evolution of physical theory.

 Two considerations govern acceptance of testable fundamental dynamical physical theories. Such a theory must be such that (i) it, together with all other accepted fundamental physical theories, exemplifies, or is a special case of, the best available metaphysical blueprint (at level 3), and physicalism (at level 4) to a sufficiently good extent, and (ii) it is sufficiently successful empirically (where empirical success is to be understood, roughly, in a Popperian sense).

 How does this hierarchical view of AOE overcome the problems and difficulties, indicated above, that confront any view which holds that science makes just one, possibly composite metaphysical assumption, at just one level? Given the one-thesis view, it must remain entirely uncertain as to what the one thesis should be. If it is relatively contentful and precise, more or less equivalent to the current level 3 thesis of AOE, then it is all too likely that this is false, and will need to be replaced in the future. If it is relatively contentless and imprecise, equivalent to theses at levels 7 or 6, this will not be sufficiently precise to exclude empirically successful but grossly *ad hoc*, aberrant theories. Even the level 4 thesis of physicalism is both too contentful and precise, and not contentful and precise enough. Physicalism may be false, and may need to be revised. At the same time, physicalism lacks the potential heuristic power to suggest good new fundamental theories which the more precise and contentful theses at level 3 possess. All these difficulties are avoided by the hierarchical view of AOE, just because of the hierarchy of assumptions, graded from the relatively contentless, imprecise and permanent at the top, to the relatively contentful, precise and impermanent (but methodologically and heuristically fruitful) at the bottom.

Any one-thesis view faces the even more serious problem of how this one thesis is to be critically assessed, revised, and improved. The hierarchical view of AOE overcomes this problem by providing severe constraints on *what* is to be revised, and *how* this revision is to proceed. In the first instance, and only in quite exceptional circumstances, only the current level 3 thesis can be revised. This revision must proceed, however, within constraints provided by the level 4 thesis of physicalism, on the one hand, and accepted, empirically successful level 2 theories, on the other hand. In a really exceptional situation, scientific progress might require the revision of the level 4 thesis of physicalism, but this too would proceed within the constraints of the thesis at level 5, and empirically successful theories at level 2, or empirically progressive research programmes at levels 2 and 3. The great merit of AOE is that it separates out what is most likely to be true from what is most likely to be false in the metaphysical assumptions of physics, and employs the former to assess critically, and to constrain, theses that fall into the latter category. It concentrates criticism and innovation where it is most likely to promote scientific progress.

Finally, any one-thesis view cannot, as we have seen, simultaneously call upon principles (1) to (4) to justify acceptance of the single thesis, whatever it may be. The hierarchical view of AOE is able to do just that. It can appeal to *different* principles, (1) to (4) above, to justify[8] (to provide a rationale for) acceptance of the *different* theses at the different levels of the hierarchy of AOE. Thus acceptance of the thesis at level 7 is justified by an appeal to (2); acceptance of theses at levels 3 to 5 are accepted as a result of (a) an appeal to (3), and (b) compatibility with the thesis above in the hierarchy. The thesis at level 6 is accepted as a result of an appeal to (4). Aberrant rivals to theses accepted at levels 3 to 5 (which might be construed to support aberrant, rival empirically progressive research programmes) are excluded on the grounds that these clash with the thesis at level (6). For further details of how AOE overcomes the two difficulties indicated above, and for further details of the view itself, see Maxwell (1998, chapter 5, and elsewhere).

It may be objected that AOE suffers from vicious circularity, in that acceptance of physical theories is justified by (in part) an appeal to physicalism, the acceptance of which is justified, in turn, by the empirical success of physical theory. My reply to this objection is that the level 6 thesis of meta-knowability asserts that the universe is such that this kind of circular methodology, there being positive feedback between metaphysics, methods, and empirically successful theories, is just what we need to employ in order to improve our knowledge. The thesis of meta-knowability, if true, justifies implementation of AOE. This response is only valid, of course, if reasons for accepting the level 6 thesis of meta-knowability do not themselves appeal to the success of science (which would just reintroduce vicious circularity at a higher level). As I made clear above, the two arguments given for accepting meta-knowability make no appeal to the success of science whatsoever.[9]

A basic idea of AOE is to channel or direct criticism so that it is as fruitful as possible,

from the standpoint of aiding progress in knowledge. The function of criticism within science is to promote scientific progress. When criticism demonstrably cannot help promote scientific progress, it becomes irrational (the idea behind (2) above). In an attempt to make criticism as fruitful as possible, we need to try to direct it at targets which are the most fruitful, the most productive, to criticize (from the standpoint of the growth of knowledge). This is the basic idea behind the hierarchy of AOE. Conjectures at all levels remain open to criticism. But, as we ascend the hierarchy, conjectures are less and less likely to be false; it is less and less likely that criticism, here, will help promote scientific knowledge. The best currently available level 3 conjecture is almost bound to be false: the history of physics reveals, at this level, as I have indicated above, that a number of different conjectures have been adopted and rejected in turn. Here, criticism, the activity of developing alternatives (compatible with physicalism) is likely to be immensely fruitful for progress in theoretical physics. Indeed, in Maxwell (1998, pp. 78-89, 159-163 and especially 217-223), I argue that this provides physics with a rational, though fallible and non-mechanical method for the discovery of new fundamental physical theories, a method invented and exploited by Einstein in discovering special and general relativity (Maxwell, 1993, pp. 275-305 ), something which Popper has argued is not possible: see Popper (1959, pp. 31-32). Criticizing physicalism, at level 4, may also be fruitful for physics, but (the conjecture of AOE is) that this is not as likely to be as fruitful as criticism at level 3. (Elsewhere I have suggested an alternative to physicalism: see Maxwell, 2005, pp. 198-205.) And, as we ascend the hierarchy (so AOE conjectures), criticism becomes progressively less and less likely to be fruitful. Against that, it must be admitted that the higher in the hierarchy we need to modify our ideas, so the more dramatic the intellectual revolution that this would bring about. If physicalism is rejected altogether, and some quite different version of the level 5 conjecture of comprehensibility is adopted instead, the whole character of natural science would change dramatically; physics, as we know it, might even cease to exist.

    The biggest change, in moving from falsificationism to AOE, has to do with the role of metaphysics in science, and the scope of scientific knowledge. According to falsificationism, untestable metaphysical theses may influence scientific research in the context of discovery, and may even lead to metaphysical research programmes; they cannot, however, be a part of scientific knowledge itself. But according to AOE, the metaphysical theses at levels 3 to 7 are all a part of current (conjectural) scientific knowledge. In particular, physicalism is. According to AOE, it is a part of current scientific knowledge that the universe is physically comprehensible - certainly not the case granted falsificationism.

    Another important change has to do with the relationship between science and the philosophy of science. Falsificationism places the study of scientific method, the philosophy of science, outside science itself, in accordance with Popper's demarcation principle. AOE, by contrast, makes scientific method and the philosophy of science an integral part of science itself. The activity of tackling problems inherent in the aims of science, at a variety of levels, and of developing new possible aims and methods, new possible more specific or less specific philosophies of science (views about what the aims and methods of science ought to be) is, according to AOE, a vital research activity of science itself. But this is also philosophy of science, being carried on within the framework of AOE.[10]

    AOE differs in many other important ways from Popper's falsificationism, whether bare or dressed (see Maxwell, 1998). Nevertheless the impulse, the intellectual aspirations and values, behind the hierarchical view of AOE are, as I have tried to indicate, thoroughly Popperian in character and spirit. The whole idea is to turn implicit assumptions into explicit conjectures in such a way that criticism may be directed at what most needs to be criticized from the standpoint of aiding progress in knowledge, so that conjectures may be developed and adopted that are the most fruitful in promoting scientific progress, at the same time no

substantial conjecture, implicit or explicit, being held immune from critical scrutiny.

6  Aim-Oriented Empiricism an Improvement over Falsificationism

AOE is also, in a number of ways, a considerable improvement over Popper's falsificationism.

1. *Consistency*.  Bare falsificationism fails dramatically to do justice to scientific practice, and is an inherently unworkable methodology, in any case.  (In what follows I shall mostly ignore bare falsificationism as obviously untenable, and concentrate on comparing dressed falsificationism and AOE.)  Dressed falsificationism does better justice to scientific practice, but at the cost of consistency; persistent rejection of empirically successful theories that do not "proceed from some simple...unifying idea" commits science to accepting a metaphysical thesis of simplicity as a part of scientific knowledge (though this is not recognized); this contradicts Popper's demarcation principle.  AOE is free of such lethal defects.

2. *Criticism*.  Pursuing physics in accordance with dressed falsificationism protects the implicit metaphysical thesis of simplicity from criticism within science itself, just because this thesis is metaphysical (and therefore not a part of science) and implicit (and therefore not available for sustained, explicit critical scrutiny).  AOE, by contrast, is specifically designed to provide a framework of metaphysical assumptions and corresponding methodological rules within which level 3 metaphysical blueprints may be developed, and critically assessed, within science.

3. *Rigou*r.  Science pursued in accordance with AOE is more rigorous than science pursued in accordance with falsificationism.  An elementary, but important requirement for rigour is that assumptions that are substantial, influential, problematic and implicit need to be made explicit so that they can be criticized, and so that alternatives can be considered.  If the attempt is made to do science in accordance with falsificationism, bare or dressed, one substantial, influential and problematic assumption must remain implicit (as we have just seen), namely the metaphysical assumption that nature behaves as if simple or unified, no *ad hoc* theory being true.  This is implicit in the adoption of the simplicity methodological principle of dressed falsificationism.  AOE, by contrast, makes this implicit assumption explicit, and provides a framework within which rival versions can be proposed and critically assessed.

4. *Simplicity*.  Falsificationism fails to say what the simplicity of a theory *is*.  Bare falsificationism provides an account of simplicity in terms of falsifiability, but we have already seen that this account is untenable.  Popper's (1963) "requirement of simplicity" appeals to a conception of simplicity or unity that is wholly in addition to falsifiability, but does not explain what the simplicity or unity of a theory is.  It fails to explain how the simplicity of a theory can possibly be methodologically or epistemologically significant when a simple theory can always be made complex by a suitable change of terminology, and *vice versa*.  Popper himself recognized the inadequacy of his simplicity requirement when he called it "a bit vague", said that "it seems difficult to formulate it very clearly", and acknowledged that it threatened to involve one in infinite regress (Popper, 1963, p. 241).  By contrast, AOE solves the problems of explaining what the simplicity or unity of a theory is without difficulty.  The totality of fundamental physical theory, T, is unified to the extent that its *content* exemplifies physicalism.  The more the content of T departs from exemplifying physicalism, the more disunified T is.[11]  Because what matters is content, not form, the way T is *formulated* is irrelevant to this way of assessing simplicity or unity.  Falsificationism cannot avail itself of this way of assessing unity because it involves acknowledging that physicalism is a basic tenet of scientific knowledge, something which falsificationism denies. Within AOE, there is a second way in which the unity of T may be assessed: in terms of the extent to which the content of T exemplifies the best available level 3 metaphysical blueprint. This second conception of simplicity or unity evolves with the evolution of level 3 ideas. As we improve our ideas about how the universe is unified, with the advance of knowledge in theoretical physics, so non-empirical methods for selecting theories on the basis of simplicity or unity improve as well.

Thus current symmetry principles of modern physics, such as Lorentz invariance and gauge invariance, which guide acceptance of theory, are an advance over simplicity criteria upheld by Newton. This account of simplicity can be extended to individual theories in two ways. First, we may treat an individual theory as a candidate theory of everything. Second, given two individual theories, $T_1$ and $T_2$, and given the rest of fundamental theory, T, $T_1$ is simpler than $T_2$ iff $T + T_1$ is simpler than $T + T_2$, where the latter is assessed in one or other of the ways indicated above.[12]

It may be objected that this proposed solution to the problem of simplicity is circular: the unity of level 2 theory is explicated in terms of the unity of level 4 physicalism. But this objection is not valid. In order to solve the problem, it is not necessary to explicate what "simplicity" or "unity" mean; rather, what needs to be done is to show how theories can be partially ordered with respect to "simplicity" or "unity" in a way that does not depend on formulation. This is achieved by partially ordering theories in terms of how well their content exemplifies the content of physicalism, so that, roughly, the more the content of a theory violates the symmetries associated with the content of physicalism, the less unity it has. As long as physicalism is a meaningful thesis, and provides a formulation-independent way of partially ordering theories in the way indicated, this suffices to solve the problem. That physicalism embodies intuitive ideas of "unity" is a bonus. For a more detailed rebuttal of this objection, see Maxwell (1998, pp. 118-123; 2005, pp. 160-174).

5. *Scientific Method*. Dressed falsificationism acknowledge (correctly) that two considerations govern selection of theory in science, namely considerations that have to do with (a) evidence, and (b) simplicity. But because it cannot solve the problem of what simplicity *is*, dressed falsificationism cannot, with any precision, specify what methods are involved when theories are selected on the basis of simplicity. Nor can the view do justice to the way in which the methods of physics evolve with evolving knowledge, especially methods that assert that acceptable theories must satisfy this or that *symmetry*. In other words, falsificationism fails to solve what may be called the "methodological" problem of induction, the problem of specifying, merely, what the methods are that are employed by science in accepting and rejecting theories (leaving aside the further problem of justifying these methods given that the aim is to acquire knowledge). AOE, by contrast, solves the problem of simplicity, and thus can specify precisely what methods are involved when theories are selected on the basis of simplicity. Furthermore, AOE can do justice to *evolving* criteria of simplicity (as we have just seen), and hence evolving methods. According to AOE, the totality of fundamental physical theory, T, can be assessed with respect to how well its content exemplifies (i) the relatively fixed level 4 thesis of physicalism, or (ii) the evolving, best available level 3 thesis. Whereas (i) constitute fixed criteria of simplicity or unity (as long as physicalism is not modified), (ii) constitute evolving criteria, criteria of unity that improve with improving knowledge.

6. *Evolving Aims and Methods*. A point, briefly alluded to in 4 and 5 above, deserves further emphasis. As physics has evolved, from Newton's time to today, non-empirical methods, determining what theories will be accepted and rejected, have evolved as well. Newton, in his *Principia*, formulated four rules of reasoning, three of which are concerned with simplicity (Newton, 1962, vol 2, pp. 398-400). Principles that have been proposed since include: invariance with respect to position, orientation, time, uniform velocity, charge conjugation, parity, time-reversal; principles of conservation of mass, momentum, angular momentum, energy, charge; Lorentz invariance; Mach's principle, the principle of equivalence; principles of gauge invariance, global and local; supersymmetry; duality principles; the principle that different kinds of particle should be reduced to one kind, and different kinds of force should be reduced to one kind; the principle that space-time on the one hand, and particles-and-forces on the other, should be unified. All of these principles can be interpreted as methodological rules which specify requirements theories must meet in order to be accepted. They can also be interpreted as physical principles, making substantial assertions about such things as space, time, matter, force. Some, such as conservation of mass, parity, and charge conjugation, have been shown to be false; others, such as Mach's principle, have never been generally accepted; still others, such as supersymmetry, remain speculative.

    Principles such as these, which can be interpreted either as physical assertions or as methodological principles, which are made explicit, developed, revised and, on occasions, rejected or refuted, are hard to account for within the framework of falsificationism. It is especially difficult, within this framework, to account for principles which (a) have a quasi *a priori* role in specifying requirements theories must satisfy in order to be accepted, but which at the same time (b) make substantial *physical* assertions about the nature of the universe. AOE, on the other hand, predicts the existence of such principles, with just the features that have been indicated. Accepted principles are components of the currently accepted level 3 blueprint. As the accepted blueprint evolves, these principles, interpreted either as physical or methodological principles, evolve as well. Indeed, according to AOE, these principles, and associated blueprints, do not just evolve; they are *improved* with improving theoretical knowledge. AOE provides a more or less fixed framework of relatively unproblematic assumptions and associated methods (at level 4 or above) within which highly problematic level 3 assumptions and associated methods may be improved in the light of the empirical success and failure of rival research programmes (which adopt rival level 3 assumptions and associated methods).

    This can be reformulated in terms of *aims* and methods of physics. A basic aim of theoretical physics is to discover the true theory of everything. This aim can be characterized in a range of ways, depending on how broadly or narrowly "theory of everything" is construed, what degree of unity such a theory must have in order to be a theory at all, and thus how much metaphysics is built into, or is presupposed by, the aim so characterized. The aim might be construed in such a way that no more than the truth of the thesis at level 7, or at level 6, is presupposed. Or, more specifically, the truth of the thesis at level 5 might be presupposed, or even more specifically, the truth of physicalism at level 4; or a range of increasingly specific blueprints at level 3 might be presupposed. Corresponding to these increasingly specific *aims* there are increasingly restrictive *methods*. As the aim becomes more specific, so it becomes more problematic, in that the presupposed metaphysics becomes increasingly likely to be false, which would make the corresponding aim unrealisable. AOE can thus be construed as providing a kind of nested framework of aims and methods, the aims becoming, as one goes down the hierarchy, increasingly problematic, and vulnerable to being unrealisable in principle, because the presupposed metaphysics is false. Within the framework of relatively unspecific, unproblematic, permanent aims and methods (high up in the hierarchy) much more specific, problematic, fallible aims and methods (low down in the hierarchy) can be revised and improved in the light of improving knowledge. There is, as I have already in effect said, something like positive feedback between improving scientific

knowledge and improving aims and methods. As knowledge improves, knowledge-about-how-to-improve-knowledge improves as well. This capacity of science to adapt itself - its aims and methods (its philosophy of science) - to what it finds out about the universe is, according to AOE, the methodological key to the astonishing progressive success of science. Falsificationism, with its fixed aim and fixed methods, is quite unable to do justice to this positive feedback, meta-methodological feature of science, this capacity of science to learn about learning as it proceeds.

7. *Verisimilitude*. The so-called problem of verisimilitude arises because physics usually proceeds from one false theory to another, thus rendering obscure what it can mean to say that science makes progress. Popper (1963, chapter 10 and Addenda) tried to solve this problem within the framework of falsificationism but, as Miller (1974) and Tichy (1974) have shown, this attempted solution does not work. Not only does falsificationism fail to specify properly the methods that make progress in theoretical physics possible; it fails even to say what progress in theoretical physics *means*.

    AOE solves the problem without difficulty. First, the fact that physics does proceed from one false theory to another, far from undermining physicalism, and hence AOE as well, is just the way theoretical physics must proceed, granted physicalism (as I have already indicated). For, granted physicalism, any theory, T*, which captures precisely how phenomena evolve in some restricted domain, must be generalizable to cover *all* phenomena. If T* cannot be so generalized then, granted physicalism, it cannot be precisely true. In so far as physics proceeds by developing theories which apply to restricted, but successively increasing, domains of phenomena, it is bound (granted physicalism) to proceed by proposing one false theory after another.

    Second, AOE solves the problem of what it can *mean* to say that theories, $T_0, ... T_N$, get successively closer and closer to the true theory-of-everything, T, as follows. For this we require that $T_N$ can be "approximately derived" from T (but not vice versa), $T_{N-1}$ can be "approximately derived" from $T_N$ (but not vice versa), and so on down to $T_0$ being "approximately derivable" from $T_1$ (but not vice versa).

    The key notion of "approximate derivation" can be indicated by considering a particular example, the "approximate derivation" of Kepler's law that planets move in ellipses around the sun (K) from Newtonian theory (NT).

    The "derivation" is done in three steps. *First*, NT is restricted to N body systems interacting by gravitation alone within some definite volume, no two bodies being closer than some given distance r. *Second*, keeping the mass of one object constant, we consider the paths followed by the other bodies as their masses tend to zero. According to NT, in the limit, these paths are precisely those specified by K for planets. In this way we recover the *form* of K from NT. *Third*, we reinterpret this "derived" version of K so that it is now taken to apply to systems like that of our solar system. (It is of course this *third* step of reinterpretation that introduces error: mutual gravitational attraction between planets, and between planets and the sun, ensure that the paths of planets, with masses greater than zero, must diverge, however slightly, from precise Keplerian orbits.)

    Quite generally, we can say that $T_{r-1}$ is "approximately derivable" from $T_r$ if and only if a theory empirically equivalent to $T_{r-1}$ can be extracted from $T_r$ by taking finitely many steps of the above type, involving (a) restricting the range of application of a theory, (b) allowing some combination of variables of a theory to tend to zero, and (c) reinterpreting a theory so that it applies to a wider range of phenomena.

    This solution to the problem of what progress in theoretical physics *means* requires AOE to be presupposed; it does not work if falsificationism is presupposed. This is because the solution requires one to assume (a) that the universe is such that a yet-to-be-discovered, true

theory of everything, T, exists, and (b) current theoretical knowledge can be approximately derived from T. Both assumptions, (a) and (b), are justified granted AOE; neither assumption is justifiable granted falsificationism.[13]

8. *Discovery of New Fundamental Theories*. Given falsificationism, the discovery of new fundamental physical theories that turn out, subsequently, to meet with great empirical success, is inexplicable. (One thinks here of Newton's discovery of his mechanical theory and theory of gravitation, Maxwell's discovery of classical electromagnetism, Einstein's discovery of the special and general theories of relativity, Bohr's discovery of "old" quantum theory, Heisenberg's and Schrödinger's discovery of "new" quantum theory, Dirac's discovery of the relativistic quantum theory of the electron and, in more recent times, the discovery of quantum electrodynamics, the electroweak theory, quantum chromodynamics and the standard model.) Granted that a new theory is required to explain a range of phenomena, there are, on the face of it, infinitely many possibilities. In the absence of rational guidance towards good conjectures, it would seem to be infinitely improbable that anyone should, in a finite time, be able to come up with a theory that successfully predicts new phenomena. The only guidance that falsificationism can provide is to think up new theories that "proceed from some simple, new, and powerful, unifying idea", in accordance with Popper's (1963) requirement of simplicity, but this is so vague and ambiguous as to be almost useless. Famously, Popper explicitly denied that a rational method of discovery is possible at all: see Popper (1959, p. 31). But if discovery is not rational, it becomes miraculous that good new theories are ever discovered. Scientific progress becomes all but inexplicable.

AOE, by contrast, provides physics with a rational, if fallible and non-mechanical, method for the discovery of new fundamental physical theories. This method involves modifying the current best level 3 blueprint so that:
(a) the new blueprint exemplifies physicalism better than its predecessor;
(b) the new blueprint promises, when made sufficiently precise to become a testable theory, to unify clashes between predecessor theories;
(c) the new theory promises to exemplify the new blueprint better than the predecessor theories exemplify the predecessor blueprint.

(a), (b) and (c) provide means for assessing how good an idea for a new theory is which do not involve empirical testing (which is brought in once the new theory has been formulated). The level 4 thesis of physicalism provides continuity between the state of knowledge before the discovery of the new theory, and the state of knowledge after this discovery. Modifying the current level 3 blueprint ensures that the new theory will be incompatible with its predecessors; it will postulate new kinds of entities, forces, space-time structure, and will exhibit new symmetries. In other words, because of the hierarchical structure of AOE, there is (across revolutions) both continuity (at level 4) and discontinuity (at levels 2 and 3), something that is not possible given falsificationism. AOE provides physics with specific non-empirical tasks to perform, specific non-empirical problems to be solved, and non-empirical methods for the assessment of ideas for new theories, all of which adds up to a rational, if fallible, method of discovery. It all stems from recognizing that physicalism is a part of current scientific knowledge. The discovery of new fundamental physical theories ceases to be inexplicable. None of this is possible granted falsificationism.[14]

The fact that AOE is able to provide a rational method of discovery, while falsificationism is not, is due to the greater rigour of AOE (a point mentioned in 3 above). AOE has greater rigour because AOE acknowledges, while falsificationism denies, metaphysical assumptions implicit in persistent scientific preference for simple, explanatory theories. It is precisely the explicit acknowledgement of these metaphysical assumptions which makes the rational method of discovery of AOE possible.

9. *Diversity of Scientific Method*.  One striking feature of natural science, often commented on, is that different branches of the natural sciences have somewhat different methods. Experimental and observational methods, and methods or principles employed in constructing and assessing theories, vary as one moves from theoretical to phenomenological physics, from physics to chemistry, from astronomy to biology, from geology to ethology. Falsificationism can hardly do justice to this striking diversity of method within the natural sciences.  Popper, indeed, tends to argue that there is unity of method, not only in natural science, but across the whole of science, including social science as well: see Popper (1961). AOE, by contrast, predicts diversity of method throughout natural science, overlaid by unity of method at a meta-methodological level.  AOE can do justice to the diversity of methods to be found in diverse sciences, without underlying unity and rationality being sacrificed.

It is important to appreciate, first, that different branches of the natural sciences are not isolated from one another: they form an interconnected whole, from theoretical physics to molecular biology, neurology and the study of animal behaviour.  Different branches of natural science, even different branches of a single science such as physics, chemistry or biology, have, at some level of specificity, different aims, and hence different methods.  But at some level of generality all these branches of natural science have a common aim, and therefore common methods: to improve knowledge and understanding of the natural world. All (more or less explicitly) put AOE into practice, but because different scientific specialities have different specific aims, at the lower end of the hierarchy of methods different specialities have somewhat different methods, even though some more general methods are common to all the sciences.  Furthermore, all natural sciences apart from theoretical physics presuppose and use results from other scientific specialities, as when chemistry presupposes atomic theory and quantum theory, and biology presupposes chemistry.  The results of one science become a part of the presuppositions of another, implicit in the aims of the other science (equivalent to the level 3 blueprint of physics, or the level 4 thesis of physicalism).  This further enhances unity throughout diversity, and helps explain the need for diversity of method.

But in order to exhibit the rationality of the diversity of method in natural science, apparent in the evolution of methods of a single science, and apparent as one moves, at a given time, from one scientific speciality to another, it is essential to adopt the meta-methodological, hierarchical standpoint of AOE, which alone enables one to depict methodological unity (high up in the hierarchy) throughout methodological diversity (low down in the hierarchy).  Falsificationism, lacking this hierarchical structure, cannot begin to do justice to this key feature of scientific method, diversity at one level, unity at another; nor can it begin to do justice to the rational *need* for this feature of scientific method.

There is a further, important point.  Any new conception of science which improves our understanding of science ought to enable us to improve scientific practice.  It would be very odd if our ability to do science well were wholly divorced from our understanding of what we are doing.  A test for a new theory of scientific method ought to be, then, that it improves scientific practice, and does not merely accurately depict current practice.  AOE passes this test.  In providing a framework for the articulation and scrutiny of level 3 metaphysical blueprints, as an integral part of science itself, thus providing a rational means for the development of new non-empirical methods, new symmetry principles, and new theories, AOE advocates, in effect, that current practice in theoretical physics be modified.  AOE makes explicit what is at present only implicit.  And more generally, in depicting scientific method in a hierarchical, meta-methodological fashion, AOE has implications for method throughout the natural sciences, and not just for theoretical physics.

In case it should seem miraculous that science has made progress without AOE being generally understood and accepted, I should add that good science has always put something close to AOE into practice in an implicit, somewhat covert way, and it is this which has made progress possible.

7 Thomas Kuhn

As I remarked in section 1 above, the main difference between Kuhn's (1962, 1970) picture of science and Popper's is that, whereas Kuhn stresses that, within normal science, paradigms are dogmatically protected from refutation, from criticism, Popper holds that theories must always be subjected to severe attempted refutation. AOE is even more Popperian than Popper's falsificationism, in that AOE exposes to criticism assumptions that falsificationism denies, and thus shields from criticism. One might think, therefore, that AOE would differ even more from Kuhn's picture of science than falsificationism does.

It is therefore rather surprising that exactly the opposite is the case. In some important respects, AOE is closer to Kuhn than to Popper.

The picture of science that emerges from Kuhn (1970) may be summarized like this. There are three stages to consider. First, there is a pre-scientific stage: the discipline is split into a number of competing schools of thought which give different answers to fundamental questions. There is debate about fundamental questions between the schools, but no overall progress, and no science.

Second, the ideas of one such school begin to meet with empirical success; these ideas become a "paradigm", and the pre-scientific school becomes normal science (competing schools withering away). Within normal science, no attempt is made to refute the paradigm (roughly, the basic theory of the science); indeed, the paradigm may be accepted even though there are well known apparent refutations. When the paradigm fails to predict some phenomenon, it is not the paradigm, but the skill of the scientist, that is put to the test. The task of the normal scientist is to solve puzzles, rather than problems. The paradigm specifies what is to count as a solution, specifies what methods are to be employed in order to obtain the solution, and guarantees that the solution exists: these are all characteristics of puzzles rather than open-ended problems. The task is gradually to extend the range of application of the paradigm to new phenomena, textbook successes being taken as models of how to proceed. Methods devolve from paradigms.

Third, the paradigm begins to accumulate serious failures of prediction; these resist all attempts at resolution, and some scientists lose faith in the capacity of the paradigm to overcome these "anomalies". A new paradigm is proposed, which does resolve these recalcitrant anomalies, but which may not, initially, successfully predict all that the old paradigm predicted. Empirical considerations do not declare that the new paradigm is, unequivocally, better than the old. Normal science gives way to a period of revolutionary science. Scientists again debate fundamentals, arguments for and against the rival paradigms often presupposing what they seek to establish. Rationality breaks down. If the revolution is successful, the new paradigm wins out, and becomes the basis for a new phase of normal science. Many old scientists do not accept the new paradigm; they die holding onto their convictions.

Kuhn argues that the dogmatic attitude inherent in normal science is necessary if science is to make progress. Applying a paradigm to new phenomena, or to old phenomena with increasing accuracy, is often extremely difficult. If every failure was interpreted as a failure of the paradigm, rather than of the scientist, paradigms would be rejected before their full range of successful application had been discovered. By refusing to reject a paradigm until the limits of its successes have been reached, scientists put themselves into a much better position to develop and apply a new paradigm. For reasons such as these, normal science, despite being ostensibly designed to discover only the expected, is actually uniquely effective in disclosing novelty.  Popper (1970), in criticizing Kuhn on normal science, ignored these arguments in support of the necessity of normal science for scientific progress.

AOE holds that much scientific work ought indeed to resemble Kuhn's normal science, in part for reasons just indicated.  But there are even more important considerations. According to AOE, and in sharp contrast with falsificationism, theoretical physics accepts a level 3 metaphysical blueprint, which exercises a powerful constraint on what kind of new theories physicists can try to develop, consider or accept.  The blueprint has a role reminiscent, in some respects, of Kuhn's paradigm, and theoretical physics, working within the constraints of the blueprint, its non-empirical methods set by the blueprint, has some features of Kuhn's normal science.

Furthermore, according to AOE, other branches of natural science less fundamental than theoretical physics invariably presuppose relevant parts of more fundamental branches.  Thus chemistry presupposes relevant parts of atomic theory and quantum theory; biology relevant parts of chemistry; astronomy relevant parts of physics.  Such presuppositions of a science have a role, for that science, that is analogous to the role that the current level 3 blueprint, or the level 4 thesis of physicalism, has for theoretical physics.  The presuppositions act as a powerful constraint on theorizing within the science. They set non-empirical methods for that science.  Such presuppositions have a role, in other words, which is similar, in important respects, to Kuhn's paradigms.  Viewed from an AOE perspective, one can readily see how and why much of science is Kuhnian puzzle-solving rather than Popperian problem-solving.

There are also, it must be emphasized, major differences between Kuhn and AOE.  The chief difference is that, according to AOE, science has a paradigm for paradigms  -  to put it in Kuhnian terms.  In order to be acceptable, level 3 blueprints must exemplify the level 4 thesis of physicalism (which in turn must exemplify the level 5 thesis of comprehensibility and so on, up to level 7).  This means that, as long as physicalism continues to be accepted as the best available level 4 thesis for science, metaphysical blueprints can be assessed in a quasi non-empirical way, in terms of how well they accord with physicalism.  Natural science is, according to AOE, one sustained, gigantic chunk of normal science, with physicalism as its paradigm.  In this respect, AOE is more Kuhnian than Kuhn (in addition to being more Popperian than Popper!).

Like falsificationism, Kuhn's picture of science is hardly tenable.  In the first place, it does not fit scientific practice very well.  Normal science undoubtedly exists, as even Popper recognized; it may well be that most scientific activity has the character of Kuhn's normal science.  But even when a discipline seems most like normal science, almost always there are a few scientists actively engaged in developing alternatives to the reigning paradigm.  And on occasions, it is from the work of these few that a new paradigm, and a new phase of normal science springs, often in a way that is quite different from Kuhn's account.  It is not obvious that accumulation of anomalies, resulting in a crisis in biology, led to Darwin's theory of evolution.  Quantum theory did not emerge, initially, from a crisis in classical physics.  Planck's work around 1900 on black body radiation engendered the quantum revolution.  It is true that classical physics, applied to a so-called black body emitting electromagnetic radiation, made a drastically incorrect prediction, but no one, not even Planck, thought that this posed a serious problem for classical physics.  The fallacious prediction of classical physics was dubbed "the ultra-violet catastrophe"; but this phrase was coined by Ehrenfest, after the quantum revolution was under way, around 1912, as propaganda for the new theory.  It was Einstein who first recognized that Planck's work spelled the downfall of classical physics; but general recognition of this only came later, probably with Bohr's quantum theory of the atom, around 1913.  Again, Einstein's general theory of relativity emerged, not because Newton's theory had accumulated anomalies and was in a state of crisis, but because it contradicted special relativity.  Einstein sought a theory of gravitation compatible with special relativity, and it was this that led him to general relativity.  These three revolutions, resulting in Darwinian theory, quantum theory and general relativity, are among the biggest and most important in the history of science; and yet they do not fit Kuhn's pattern.

Failure to fit scientific practice in detail does not, however, provide decisive grounds for rejecting a normative account of scientific method.  One can always reply that the account specifies how science *ought* to proceed, not how it has in fact proceeded.  Much more serious are the objections of principle to Kuhn's account.  Kuhn, like Popper, provides no account of the creation of new paradigms.  And given Kuhn's insistence that a new paradigm, after a successful revolution, is  incommensurable with its pre-revolutionary predecessor, it would seem impossible to provide rational (if fallible) procedures for the creation of good new paradigms while maintaining consistency with the rest of Kuhn's views.  Kuhn does allow that non-empirical criteria, or values, such as consistency and simplicity, are employed by science permanently (and therefore, presumably, across revolutions) to assess theories or paradigms; but Kuhn also emphasizes that these criteria are flexible, and open to different interpretations (Kuhn, 1970, p. 155; 1977, ch. 13).  There is no account of what simplicity is, and no advance over Popper's "requirement of simplicity".  Furthermore, Kuhn's appeal to simplicity faces the same difficulty we have seem arising in connection with Popper's appeal to simplicity.  If "simplicity" is interpreted in such a way that it has real content, and is capable of excluding "complex" or disunified and aberrant theories or paradigms from science, then its permanent employment by science commits science to a permanent metaphysical assumption that persists through revolutions, something Kuhn explicitly rejects (and could not, in any case, provide a rationale for).  If "simplicity" is interpreted sufficiently loosely and flexibly to ensure that no such metaphysical thesis is involved, invoking simplicity must fail to exclude complex, disunified, aberrant paradigms from science.  Any Kuhnian requirement of simplicity, in short, must either be incompatible with the rest of Kuhn's views, or toothless and without content.  Either way, Kuhn has no consistent method for excluding complex, aberrant paradigms from consideration.  It should be noted that Kuhn is emphatic that no sense can be made of the idea that there is progress in knowledge across revolutions, the new paradigm being better, closer to the truth, than the old one: see Kuhn (1970, chapter XIII).  But this is a disaster for Kuhn's whole view. Why engage in normal science if the end result is the rejection of all that has been achieved, all the progress in knowledge of that period of normal science being sacrificed when the science adopts a new paradigm?  Kuhn's arguments for the progressive character of normal science, indicated above, are all defeated.

Perhaps the most serious objection to Kuhn's picture of science is the obvious basic unintelligence of its prescriptions for scientific research.  Suppose we have the task of crossing on foot difficult terrain, containing ravines, cliffs, rivers, swamps, thickets.  Kuhn's view, applied to this task, would be as follows.  After debate about which route to follow (pre-science), one particular route is chosen and then followed with head down, no further consideration being given to changing the route (normal science).  Eventually, this leads to an impasse: one comes face to face with an unclimbable cliff, or finds oneself waist deep in a swamp, and in danger of drowning (crisis).  Finding oneself in these dire circumstances, a new route is taken (new paradigm), and again, with head down, this new route is blindly followed (normal science) until, again, one finds oneself unable to proceed, about to drown in a river, or tumble into a ravine.

This is clearly a stupid way to proceed.  It would be rather more intelligent if, as one tackles immediate problems of wading through this stream, climbing down this scree (puzzle-solving of normal science), one looks ahead, whenever possible, and reconsiders, in the light of the terrain that has been crossed, what adjustments one needs to make to the route one has opted to follow.  Exactly the same point holds for science.  There can be division of labour.

Even if a majority of scientists tackle the multitude of puzzles that go to make up normal scientific research, taking the current theory, or paradigm, for granted, there ought also to be some scientists who are concerned to look ahead, consider more fundamental problems, explore alternatives to the current paradigm.  In this way new paradigms may be developed before  science plunges deep into crisis.  And just this does go on in scientific practice, as I have already indicated in the brief discussion of the work of Darwin and Einstein (and somewhat less convincingly, Planck).  Another example of a new, revolutionary theory or paradigm being proposed in the absence of crisis is Wegener's advocacy of the movement of continents, anticipating the plate tectonic revolution by decades.  Science is, in practice, more intelligent than Kuhn allows.

In sharp contrast to Kuhn, AOE does not merely stress the importance of "looking ahead", of trying to develop new theories, new paradigms before science has plunged into crisis; even more important, AOE provides a framework for theoretical physics (and therefore, in a sense, for the whole of natural science) within which ideas for fundamental new theories may be developed and assessed.

According to Kuhn, successful revolutions mark radical discontinuities in the advancement of science, to the extent, indeed, that old and new paradigms are "incommensurable" (i.e. so different that they cannot be compared).  This Kuhnian view is most likely to be correct when applied to revolutions in fundamental theoretical physics, where radical discontinuity seems most marked.  But it is precisely here that Kuhn's claim turns out to be seriously inadequate.  As I have already emphasized, all revolutions in theoretical physics, despite their diversity in other respects, reveal one common theme: they are all gigantic steps in unification.  From Newton, via Maxwell, Einstein, Bohr, Schrödinger and Dirac, to Salam, Weinberg and Gell-Mann, all new revolutionary theories in physics bring greater unity to physics.  (And Darwinian theory, one might add, brings a kind of unity to the whole of biology.)  The very phenomenon that Kuhn holds to mark discontinuity, namely revolution, actually also reveals continuity  - continuity of the search for, and the successful discovery of, underlying theoretical unity.

This aspect of natural science, to which Kuhn fails entirely to do justice, is especially emphasized by AOE.  According to AOE, revolutions in theoretical physics mark discontinuity at the level of theory, at level 2, and even discontinuity at level 3, but continuity at level 4.  Physicalism, which asserts that underlying dynamic unity exists in nature, persists through revolutions  - or, at least, has persisted through all revolutions in physics since Galileo.  In order to make rational sense of natural science, we need to interpret the whole enterprise as seeking to turn physicalism, the assertion of underlying dynamic unity in nature, into a precise, unified, testable, physical "theory of everything".  That, in a sentence, is what AOE asserts.  Physicalism, according to AOE, despite its metaphysical (untestable) character, is the most secure item of theoretical knowledge in science; it is *the* most fruitful idea that science has come up with, at that level in the hierarchy of assumptions.

Because of its recognition that, despite the discontinuity of revolutions at levels 2 and 3, there is the continuity of the persistence of physicalism at level 4 (and of other theses at levels higher up in the hierarchy), AOE is able to resolve problems concerning the discovery and assessment of paradigms which Kuhn's view is quite unable to solve.  Both fundamental physical theories, and level 3 blueprints, can be partially ordered with respect to how well they exemplify physicalism, entirely independent of ordinary empirical assessment.  Assessing progress through revolution poses no problem for AOE.  As we have seen, AOE solves the problem of verisimilitude.

I have already mentioned that AOE does not merely describe scientific practice; it carries implications as to how scientific practice can be improved.  One such implication concerns scientific revolutions.  Kuhn (1970) gives a brilliant description of the way, during a scientific revolution, there is a breakdown of rationality, competing arguments for the rival paradigms being circular, each presupposing what is being argued for.  This is a feature of actual science.  Scientists do not know how to assess competing theories objectively, when empirical considerations are inconclusive.  But all this can be seen to be a direct consequence of trying to do science without persisting metaphysical assumptions concerning the comprehensibility of the universe, there thus being nothing available to constrain acceptance of theories when empirical considerations are inconclusive.  Consider Kuhn's breakdown of rationality.  A substantial revolution will involve, not just two rival paradigms or theories, $T_1$ and $T_2$, but two rival blueprints, $B_1$ lurking behind $T_1$, and $B_2$ lurking behind $B_2$.  Granted $B_1$, $T_1$ is far more acceptable than $T_2$, but the reverse granted $B_2$.  But $B_1$ and $B_2$, being untestable, metaphysical theses, are not explicitly discussable, and objectively assessable, within science: so they are more or less repressed, excluded from discussion.  Nevertheless, scientists do think in terms of $B_1$ and $B_2$.  Kuhn's Gestalt switch, involved in switching allegiance from $T_1$ to $T_2$, can be pin-pointed as the act of abandoning the old blueprint and adopting the new one.  Non-empirical arguments in favour of $T_1$ or $T_2$ can only take the form of an appeal to $B_1$ or $B_2$, in however a muffled a way (due to the point that blueprints are not open to explicit discussion).  Such arguments will be circular, and entirely unconvincing to the opposition, in just the way described by Kuhn.  Accept $B_1$, and $T_1$ becomes the only possible choice; accept $B_2$ and $T_2$ is the only choice.  Each side in the dispute is convinced that the other side is wrong, even incoherent.  What needs to be done, and cannot be done, of course, is to discuss the relative merits of $B_1$ and $B_2$.  Just this can be done, granted AOE.  $T_1$, $B_1$, $T_2$ and $B_2$ can all be assessed from the standpoint of adequacy in exemplifying physicalism.  When the scientific community adopts AOE, the Kuhnian irrationality of revolutions will disappear from science.

It may be asked: How is it possible for AOE to be *both* more Popperian than Popper, and more Kuhnian than Kuhn?  The answer is that AOE is more Popperian that Popper in making explicit, and so criticizable, metaphysical theses which falsificationism denies, and thus leaves implicit and uncriticizable within science.  But AOE is also more Popperian than Popper in insisting we need to exploit criticism *critically*, so that it furthers, and does not sabotage, the growth of knowledge.  Criticism needs to be marshalled and directed at that part of our conjectural knowledge which it is, we conjecture, the most fruitful to criticize.  This means directing critical fire at level 2 theories and level 3 blueprints, it being less likely, though still possible, that criticism of the level 4 thesis of physicalism will aid the growth of empirical knowledge.  Physicalism has played an extraordinarily fruitful role in the advancement of scientific knowledge; it should not be abandoned unless an even more apparently fruitful idea is forthcoming, or unless the empirical and explanatory success that physicalism appears to have engendered turns out to be illusory.  It is this persistence of physicalism, for good Popperian reasons, which gives to theoretical physics, and indeed to the whole of natural science, something of the character of Kuhn's normal science, with physicalism as its quasi-permanent "paradigm".

8 Imre Lakatos

Lakatos sought to reconcile the very different views of science held by Popper and Kuhn.  According to Kuhn, far from seeking falsifications of the best available theory, as Popper held, scientists protect the accepted theory, or "paradigm", from refutation for most of the time, the task being to fit recalcitrant phenomena into the framework of the paradigm.  Only when refutations become overwhelming, does crisis set in; a new paradigm is sought for and found, a revolution occurs, and scientists return to doing "normal science", to the task of reconciling recalcitrant phenomena with the new paradigm.  Lakatos sought to reconcile Popper and Kuhn by arguing that science consists of competing fragments of Kuhnian normal science, or "research programmes", to be assessed, eventually, in terms of their relative empirical success and failure. Instead of research programmes running in series, one after the other, as Kuhn thought, research programmes run in parallel, in competition, this doing justice to Popper's demand that there should be competition between theories (a point emphasized especially by Feyerabend).[15] Lakatos became so impressed with the Kuhnian point that theories always face refutations, the empirical successes of a theory being a far more important guide to scientific progress than refutation, that he finally came to the conclusion that Popper's philosophy of science was untenable.

AOE has a number of features in common with Lakatos's methodology of scientific research programmes. AOE makes extensive use of the notion of scientific research programme. Like Lakatos's view, AOE exploits the idea that such research programmes can, sometimes, be compared with respect to how empirically progressive they are. AOE, again like Lakatos's view, sees the whole of science as a gigantic scientific research programme. And like Lakatos's view, AOE can be construed as synthesizing Popper's and Kuhn's views.

But there are also striking differences. There are differences in the way scientific research programmes are conceived, especially research programmes in fundamental physics. For Lakatos, main components of a research programme are the "hard core" (corresponding to Kuhn's "paradigm"), and the "protective belt" of "auxiliary hypotheses", which facilitate the application of the hard core to empirical phenomena. The main business of a research programme is to develop the protective belt, thus extending, and making more accurate, the empirical predictions of the hard core. The hard core is a testable theory rendered metaphysical by the methodological decision not to allow it to be refuted, refutations being directed at the protective belt rather than the hard core.

According to AOE, by contrast, the metaphysical kernel of a research programme is not a testable theory but rather a thesis that is genuinely metaphysical (i.e. more or less unspecific, and usually untestable) - a thesis such as the corpuscular hypothesis, Boscovich's point-atom blueprint, Einstein's unified field blueprint, and so on. The basic aim of the programme is to turn the relatively unspecific blueprint into a precise, testable (and true) physical theory. The research programme thus consists of a succession of theories, $T_1$, $T_2$,...$T_n$, which can be compared, not only with respect to empirical success, but also with respect to how adequately each theory encapsulates, or exemplifies, the blueprint of the programme. (The latter is not possible within a Lakatosian programme.) Whereas a Lakatosian programme has a fixed basic theory (or hard core), and seeks to improve auxiliary hypotheses (the protective belt), an AOE programme strives to capture the blueprint more and more adequately by means of testable physical theories.

Both Lakatos's view and AOE permit one to see natural science as one gigantic research programme, but how this programme is construed is very different. For Lakatos "science as a whole can be regarded as a huge research programme with Popper's supreme heuristic rule: 'devise conjectures with more empirical content than their predecessors'" (1970, p. 132). The huge research programme of natural science has, for Lakatos, no hard core; to this extent, Lakatos's view is a variant of Popper's.[16] According to AOE, however, if natural science is viewed as one gigantic research programme, then it does have something like a hard core. First, there is physicalism at level 4, a metaphysical but nevertheless substantial thesis about the nature of the universe. And then there is the current blueprint at level 3, an even more substantial metaphysical thesis about the nature of the universe. These provide severe constraints on what theories are acceptable that are not straightforwardly empirical,[17] something that is not possible given the views of Popper or Lakatos[18] (or even Kuhn).

Lakatos and AOE have very different motivations for taking scientific research programmes so seriously. For Lakatos, the motivation comes from appreciating that a scientific theory, T, cannot be decisively refuted at an instant, as it were, partly because auxiliary hypotheses can always be invented to salvage T from a refutation, partly because early applications of a new theory, such as Newton's, may make simplifying assumptions which may well lead to false predictions (the fault lying with the simplifying, auxiliary hypotheses rather than the basic theory). Only by looking at a series of theories, a given $T_1$ (the hard core) plus changing auxiliary hypotheses (the protective belt), and comparing this with a rival series based on a different hard core, $T_2$, and comparing the extent to which the two series are empirically progressive or degenerating, can one assess the relative empirical merits of $T_1$ and $T_2$. For AOE, the situation is very different. A research programme in theoretical physics consists of a blueprint, B, and a succession of theories, $T_1, T_2...T_n$ (each equivalent to a Lakatosian hard core), successive attempts to capture B as a testable theory. If $T_1, T_2...T_n$ are increasingly empirically successful (in a roughly Popperian sense) and also increasingly successful at capturing B, then this means that B is empirically fruitful. A rival blueprint, B*, might be such that the series $T_1, T_2...T_n$ moves further and further away from B*: this would mean that B* is empirically sterile. A major part of the point of research programmes, for AOE, is to assess the relative empirical fruitfulness of rival metaphysical theses, at levels 3 and 4 (and above, if necessary). Though mostly untestable, nevertheless metaphysical theses can be assessed in a quasi-empirical way, in terms of the empirical progressiveness or degeneracy of the research programmes with which they are associated (or can be regarded as being associated).[19] This is, according to AOE, a key feature of scientific method, one which makes scientific progress possible. It makes it possible for improving theoretical knowledge to lead to a reassessment of what is the best available blueprint, which in turn leads to a reassessment of the best available non-empirical methodological rules, such as symmetry principles. In other words, it makes it possible for there to be positive feedback between improving knowledge and improving aims-and-methods (improving knowledge-about-how-to-improve-knowledge), a vital feature of scientific rationality according to AOE.

The differences indicated enable AOE to overcome problems which Lakatos's view cannot solve. Lakatos insists that there is no such thing as instant rationality: however apparently decisive the refutation of a theory may be, it is always possible to salvage it from refutation in a content increasing way by the invention of an appropriate auxiliary hypothesis. It is this consideration which leads Lakatos to argue that only series of theories, competing research programmes, can be assessed rationally, in terms of relative empirical progressiveness. But in practice in science there do seem to be instant refutations. A famous example is the refutation of parity. This is a symmetry which declares, roughly, that if a process can occur, then so can its mirror image. This was decisively refuted by Wu et al. (1957), by means of an experiment which showed that electrons were emitted in a preferential direction from cobalt nuclei undergoing radioactive decay in a magnetic field. Parity conservation implied that this would not occur. Strictly speaking, it was not parity conservation on its own that was refuted, but parity plus quantum theory plus the theory of weak interactions plus the theory of nuclear structure plus a highly theoretical description of the experiment. One would think there was plenty of scope, here, for auxiliary hypotheses to be invented to salvage parity from refutation. No such hypothesis was forthcoming; the refutation of parity conservation was accepted immediately by the physics community, despite strong resistance to accepting such a conclusion (because of the implausibility of supposing that nature distinguishes between left-handedness and right-handedness at the level of fundamental physical theory). Allan Franklin, who has produced what is probably the best account of the downfall of parity conservation, has put the matter like this: "It is fair to say that as soon as any physicist saw the experimental result they were convinced that parity was not conserved in the weak interactions" (Franklin, 1990, p. 66).[20] Scientific practice seems almost to refute Lakatos's view.

But it does not refute AOE. According to Lakatos, in the end only empirical considerations, plus considerations of empirical content, restrict choice of theory; few restrictions are placed on how a body of theory may be modified to salvage it from refutation. AOE places much more severe restrictions on choice of theory. In addition to those that it has in common with Lakatos's view, AOE demands of a fundamental physical theory that it, together with other such theories, exemplifies physicalism, to a sufficient degree. This makes it very much more difficult to modify a body of theory so as to salvage it from refutation. Instant refutation is not surprising, granted AOE.

Lakatos's view requires that science consists of competing research programmes. Unquestionably, the history of science reveals that competing research programmes have, on occasions existed. But it is not clear that all science has this character, as Lakatos's view would seem to require. After Heisenberg and Schrödinger had developed quantum theory in the mid 1920's, there continued to be debate about how the new theory should be interpreted, and whether the new theory, interpreted along the orthodox lines advocated by Bohr, Heisenberg and others, was ultimately acceptable. But there was nothing like a competing research programme. Viewed from the perspective of AOE, all this makes perfect sense. There were indeed serious grounds for regarding the new theory as unsatisfactory (see Maxwell, 1998, chapter 7). But the new theory had achieved such striking successes, it was rational to conjecture that progress lay in developing the new theory, applying it to new phenomena, reconciling it with special relativity - in doing something like Kuhnian normal science, in other words - rather than in trying to develop a rival theory, a rival research programme. (To say this is not to say that serious attention should not have been given to the theoretical defects of orthodox quantum theory.) Not only does the history of science fail to reveal that there are always competing research programmes; whenever a new theory arrives on the scene that meets with extraordinary empirical success and no refutation, no good rationale may exist for inventing a rival research programme. (As we have seen, unlike Popper's falsificationism and Lakatos's research programme view, AOE holds that something like Kuhn's normal science may well be rational, as long as it is accompanied by some sustained tackling of problems associated with the currently accepted blueprint. This may, eventually, but not immediately, lead to the development of a new fundamental theory, a new research programme.)

There are other, much more decisive ways in which AOE is an improvement over Lakatos's view. Lakatos's methodology of research programmes inherits a number of unsolved problems from its two sources, Popper and Kuhn. Like Popper and Kuhn, Lakatos has no solution to the problem of what the simplicity, unity or explanatory character of a theory, or hard core, is; AOE, as I have indicated briefly above, solves the problem without difficulty. In failing to say what simplicity *is*, Lakatos also fails to articulate with any precision that part of scientific method concerned with simplicity; AOE faces no difficulty here either. Like Popper and Kuhn, Lakatos can say nothing useful about how new theories, new hard cores, are created or discovered; AOE, as a result of including levels 3 and 4 within the domain of scientific knowledge, is able to specify a rational, if fallible and non-mechanical, method for the creation of new theories, even new fundamental theories of physics. Finally, Lakatos's view fails to solve the problem of verisimilitude, a problem which can be readily solved granted AOE.

Popper, Kuhn and Lakatos, despite their differences, have one big failure in common (the source of almost all the others). All three take for granted that:
(A) In science no untestable but nevertheless substantial thesis about the world can be accepted as a part of scientific knowledge in such a firm way that theories which clash with it, even if highly successful empirically, are nevertheless rejected.

Popper accepts (A) in that, for him, untestable theses are metaphysical, and therefore not a part of scientific knowledge. Kuhn holds it, because, for Kuhn, nothing theoretical survives a revolution. Kuhn's acceptance of (A) is also apparent in his whole treatment of revolutions: precisely because Kuhn accepts (A), Kuhn cannot invoke anything like the level 4 thesis of physicalism to assess rival paradigms during a revolution, when empirical considerations are inconclusive. The Kuhnian irrationality of revolutions is a consequence of scientists accepting (A); and in so far as Kuhn thinks this irrationality is inevitable, Kuhn accepts (A) as well.

A case could be made out for saying that Lakatos came near to rejecting (A) in arguing for the need for science to adopt a conjectural metaphysical inductive principle which, if true, would more or less guarantee that Popperian, or rather Lakatosian, methods deliver authentic theoretical knowledge.

But Lakatos here missed the fundamental point, central to AOE, and highly Popperian in spirit, that our current methods are all too likely to be more or less the *wrong* methods to adopt, the metaphysics implicit in these methods being *false*, there thus being a vital need, for scientific progress, to make the metaphysics explicit so that it can be criticized, so that alternatives can be developed and considered, leading to *improved* metaphysics and methods, this in turn requiring the development of a hierarchy of metaphysical theses to form a framework of relatively unproblematic theses within which more specific problematic theses may be developed and assessed.

Interestingly enough, Lakatos himself was aware of this deficiency in his "plea to Popper for a whiff of 'inductivism'" (1978, p. 159). Discussing his proposal that one should appeal to a metaphysical inductive principle as a conjecture as a part of the solution to the problem of induction, Lakatos says:

"Alas, a solution is interesting only if it is embedded in, or leads to, a major research programme; if it creates new problems - and solutions - in turn. *But this would be the case only if such an inductive principle could be sufficiently richly formulated so that one may, say criticize our scientific game from its point of view*. My inductive principle tries to explain why we 'play' the game of science. But it does so in an *ad hoc*, not in a 'fact-correcting (or, if you wish, 'basic value judgment correcting') way" (Lakatos, 1978, p. 164).

Lakatos highlights, here, the difference between his own position and that of AOE. The (revisable) AOE thesis of physicalism is indeed "sufficiently richly formulated so that one may...criticize our scientific game from its point of view". AOE not only offers a new research programme for the philosophy of science; it modifies the research programme of science, one modification being that the philosophy of science becomes an integral part of science itself. The passage above makes me wonder whether Lakatos might not have gone on to develop or endorse AOE if he had lived.

Nicholas Maxwell
University College London

Notes
1. The version of AOE defended here is a simplification and improvement of the version expounded in Maxwell (1998), in turn an improvement of versions of the view expounded in Maxwell (1972, 1974, 1979, 1984, 1993 and 1997). For summaries of (1998) see Maxwell (1999, 2000, 2002a, 2002b).
2. See Popper (1959, 1963, 1983)
3. See Lakatos (1970, 1978).
4. For Popper's replies to such criticisms: see Popper (1972), chapter 1; (1974), sections II and III; and (1983), Introduction and chapter 1.
5. Popper discusses such "silly" rival theories in Popper (1983, pp. 67-71). He argues that they deserve to be rejected on the grounds that they create more problems than they solve, problems of explanation. This is a relevant consideration granted dressed falsificationism, but not granted bare falsificationism. He also argues that it does not matter if such "silly" theories become potential rivals, since it can be left to scientists themselves to criticize them. But what this ignores is that it is precisely Popper's methodology which should be providing guidelines for such criticism. Far from condemning such a "silly" theory as worthy of rejecting, bare falsificationism holds such a theory to be better than the accepted theory (if it has greater empirical content, is not falsified where the accepted theory appears to be, and some of the excess content of the "silly" theory is corroborated). Popper fails to appreciate that it his methodology, not he himself, which needs to declare that silly theories are indeed "silly". The fact that his methodology declares these silly theories to be highly acceptable is a devastating indictment of his methodology. To argue that these silly theories, refuting instances of his methodology, do not matter and can be discounted, is all too close to a scientist arguing that evidence, that refutes his theory, should be discounted, something which Popper resoundingly condemns. The falsificationist stricture that scientists should not discount falsifying instances, ought to apply to methodologists as well!
6. In fact even the methodological rules of bare falsificationism are such that persistent application of these rules commits one to making implicit metaphysical assumptions (which may be false). Bare falsificationism, as formulated by Popper, requires of an acceptable theory that it is strictly universal in that it makes no reference to any specific time, place or object. This makes it impossible for science to discover that the laws of nature just are different within specific space-time regions, or that there is a specific object with unique dynamical properties. There is no scope, within bare falsificationism, for the rejection of these metaphysical theses, even though circumstances could conceivably arise such that progress in knowledge would require this. (AOE, by contrast, allows for this remote possibility: that which is dogmatically upheld by bare falsificationism becomes criticizable granted AOE.) Popper recognizes that the methodological rule requiring any theory to be strictly universal does have a metaphysical counterpart (1959, sections 11 and 79), but fails to appreciate how damaging this is for falsificationism.

7. Smart (1963) has used the term 'physicalism' to stand for the view that the world is made up entirely of physical entities of the kind postulated by fundamental physical theories - electrons, quarks and so on. As I am using the term, 'physicalism' stands for the very much stronger doctrine that the universe is physically comprehensible, that it is such that some yet-to-be-discovered, unified "theory of everything" is true.

8. This talk of "justifying" may seem thoroughly unPopperian in character, but it is not. What is at issue is not the justification of the truth, or probable truth, of some thesis, but only the justification of *accepting* the thesis (granted our aim is truth). Within Popper's falsificationism, there is just such a "justification" for accepting highly falsifiable (and unfalsified) theories: such theories, being most vulnerable to falsification, facilitate the discovery of error, and thus give the most hope of progress (towards truth). Acceptance of such theories is justified (according to falsificationism) because it promotes error detection and progress.

9. For a more detailed rebuttal of this objection see Maxwell (2005, pp. 207-210).

10. In holding that metaphysical theses and philosophies of science are an integral part of science itself, AOE implies that Popper's principle of demarcation (Popper, 1963, chapter 11) is to be rejected. Popper's demarcation proposal, apart from being untenable, is in any case too simplistic, in that it reduces to one a number of distinct demarcation issues. Popper rolls into one the distinct tasks of demarcating (a) good from bad science, (b) science from non-science, (c) science from pseudo-science, (d) rational from irrational inquiry, (e) knowledge from mere speculation, (f) knowledge from dogma (or superstition, or prejudice, or popular belief), (g) the empirical from the metaphysical, and (h) factual truth from non-factual (analytic) truth. (a) to (d) involve demarcating between disciplines, whereas (e) to (h) involve demarcating between propositions.

11. Dynamical theories are partially ordered with respect to the extent that they exemplify physicalism, with respect to their degree of unity, in other words. For further details see Maxwell (1998, chapter 4).

12. For a very much more detailed exposition of this solution of the problem of simplicity, together with an account of the way in which great unifying theories of physics illustrate the solution, see Maxwell (1998, chapters 3 and 4). See also Maxwell (2005, pp. 160-174).

13. It may be objected that if T is assumed to be the true *unified* theory of everything, no meaning can be given to the idea that theoretical physics is making progress, by means of a succession of false theories, to a more or less *disunified* theory of everything. But T does not need to be assumed to be unified; all that is required is that T is such that the notion of "partial derivation" from T makes sense. For further discussion of the inability of any standard empiricist view such as falsificationism to solve the problem of verisimilitude, and the ability to AOE to solve the problem, see Maxwell (1998, pp. 70-72, 211-217 and 226-227).

14. For further discussion of the method of discovery provided by AOE see Maxwell (1974,

Part II; (1993, Part III); and (1998, pp. 159-163 and 219-223).

15. See Lakatos (1970, 1978). For Feyerabend's argument that severe testing requires the development of rival theories see Feyerabend (1965).

16. Granted Lakatos's overall view, the research programme of science cannot have a hard core, for then, in order to ensure Popperian severe testing, there would need to be a rival research programme with a rival hard core - and that would mean the original research programme was not the whole of science. Actually, Lakatos is not quite consistent here; after the sentence quoted in the text, Lakatos goes on "Such methodological rules may be formulated, as Popper has pointed out, as metaphysical principles. For instance, the *universal* anti-conventionalist rule against exception-barring may be stated as the metaphysical principle: 'Nature does not allow exceptions'" (1970, p. 132). That this admission is damaging for Popper's bare falsificationism was pointed out in footnote 6; it is equally damaging for Lakatos's version of Popperianism.

17. I say "not straightforwardly empirical" because both physicalism and the best available blueprint are themselves accepted on the grounds that they support a more empirically progressive research programme than any rival theses. Long-term empirical considerations influence choice of theses at levels 3 and 4, while at the same time these theses can lead to the rejection of potentially empirically successful theories that clash too severely with them (i.e. are too severely *ad hoc*).

18. The Popperian and Lakatosian demand that theories be strictly universal places weak but rigid constraints on what theories are acceptable; the demand of AOE that theories accord, as far as possible, with physicalism and the best available blueprint, places strong, but flexible and revisable constraints on what theories are acceptable. For further discussion see Maxwell (1998, pp. 89-102, chapter 4, and 223-227.)

19. For further details and discussion, see Maxwell (1998, pp. 172-180).

20. For an account of the discovery of parity non-conservation, and of the decisive character of the experiments refuting parity conservation, see Franklin (1990, pp. 63-6 and 151-2). See also Franklin (1986).